\documentclass[12pt]{article}
\usepackage{amsmath}
\usepackage{amssymb}
\usepackage{latexsym}
\usepackage{graphicx}
\usepackage{psfrag,fancyhdr,epsfig}
\def\beq{\begin{equation}}
\def\eeq{\end{equation}}
\def\bea{\begin{eqnarray}}
\def\eea{\end{eqnarray}}

\begin{document}

\begin{titlepage}

\vspace*{1cm}

\begin{center}
{\bf {\Large {Greybody factors in a rotating black-hole
background-II : fermions and gauge bosons.}}}

\bigskip \bigskip \medskip

{\bf S. Creek}$^1$, {\bf O. Efthimiou}$^2$, {\bf P. Kanti}$^{1,2}$
and {\bf K. Tamvakis}$^2$

\bigskip
$^1$ {\it Department of Mathematical Sciences, University of Durham,\\
Science Site, South Road, Durham DH1 3LE, United Kingdom}

$^2$ {\it Division of Theoretical Physics, Department of Physics,\\
University of Ioannina, Ioannina GR-45110, Greece}

\bigskip \medskip
{ \bf{Abstract}}
\end{center}

We study the emission of fermion and gauge boson degrees of
freedom on the brane by a rotating higher-dimensional black hole.
Using matching techniques, for the near-horizon and far-field
regime solutions, we solve analytically the corresponding field
equations of motion. From this, we derive analytical results for
the absorption probabilities and Hawking radiation emission rates,
in the low-energy and low-rotation case, for both species of
fields. We produce plots of these, comparing them to existing
exact numerical results with very good agreement. We also study
the total absorption cross-section and demonstrate that, as in the
non-rotating case, it has a different behaviour for fermions and
gauge bosons in the low-energy limit, while it follows a universal
behaviour -- reaching a constant, spin-independent, asymptotic
value -- in the high-energy regime.

\end{titlepage}

\setcounter{page}{1}
\noindent

\section{Introduction}
The leading motivation for studying higher-dimensional theories
\cite{ADD,RS} is that they provide a framework for the unification
of gravitation with the rest of the fundamental forces. Gravity, and
possibly scalar fields, propagate in a $(4+n)$-dimensional spacetime
(the \textit{Bulk}), while ordinary matter is confined in a
four-dimensional hypersurface, the \textit{Brane}. In models with
large extra dimensions, the traditional Planck scale is an effective
scale, while the fundamental Planck scale of gravitation is related
to it in terms of the number and size of the extra dimensions.
In such a framework higher-dimensional black holes can be created
in trans-planckian collisions
\cite{creation}. This opens the possibility of seeing them in
ground-based colliders \cite{colliders}, or in cosmic ray interactions
\cite{cosmic}. These higher-dimensional black holes and their
properties have been the subject of a number of articles in the
last few years -- for a summary of their properties and phenomenological
implications, see \cite{ADMR,Kanti2004, reviews}.

A black hole created in a
high energy collision is expected to evaporate through Hawking
radiation \cite{hawking} both in the bulk, through the emission of
gravitons and scalar fields, and on the brane through the emission
of fermions and gauge bosons. The black hole is expected to
undergo a number of phases: First, the
\textit{balding} phase, in which the black hole emits mainly
gravitational radiation, losing all ``hair'' inherited from the
original particles. Then, in the \textit{spin-down} phase, the
black hole loses all its angular momentum, through the emission of
Hawking radiation. Third is the \textit{Schwarzschild} phase, in
which the black hole loses its actual mass due to the emission of Hawking
radiation. Finally, the \textit{Planck} phase -- where the black
hole's mass or temperature reach the characteristic scale of
gravity -- which needs a quantum gravity theory in order to be
studied. There have been both numerical and analytical studies of
the Hawking radiation emitted from such a higher-dimensional black
hole, for the Schwarzschild phase \cite{kmr1, FS, HK1,
Barrau, Jung, BGK, Naylor, Park, Cardoso, Creek, Dai, Liu, Chen}, as well as
for the spin-down phase \cite{Frolov2, IOP1, HK2, IOP-proc, DHKW, CKW, CDKW, IOP2,
Jung-super, rot-bulk, Jung-rot, CEKT2}.

In the present work we focus on the evaporation of a rotating
$(4+n)$-dimensional black hole through Hawking radiation in the
form of fermions and gauge bosons on the brane. Numerical studies
of fermion and gauge boson emission by a rotating black hole on
the brane already exist in the literature \cite{CKW, CDKW, IOP2}.
In this article, we continue our effort, initiated in \cite{CEKT2}
with the study of scalar fields, to derive complementary analytic
results for higher-spin fields using an analytical approach. In
section 2, we consider the gravitational background corresponding
to a higher-dimensional rotating black hole, and write down the
equations, describing all radiative components of fermion and gauge
boson fields propagating on the brane, in the form of a single master
equation. In section 3, we solve this equation analytically by
using a well-known solution matching technique: we first derive the
solution at the near horizon regime, and then at the far field regime.
Next, we stretch and match the two solutions at the intermediate zone,
in the low-energy and low-rotation limit, thus producing a smooth
solution for all spacetime. We then use this solution in order to
compute, for every type of field, the absorption probability, a quantity
that characterizes the Hawking radiation. In section 4 we produce plots
for the absorption coefficient and compare them with existing numerical
results. In sections 5 and 6, we study the asymptotic behaviour of
the corresponding cross-section and the profile of the energy emission
rates, respectively. In section 7, we state our conclusions.


\section{Master equation on a brane embedded in a rotating $(4+n)D$
black hole background}

The background around a $(4+n)$-dimensional rotating black hole is
given by the Myers-Perry solution \cite{MP}. As is usual, here we focus
on the case where the black hole, being created by the collision of
brane-localised particles, is characterised by only one non-zero
angular momentum component parallel to our brane. Then, the
line-element takes the form
\begin{eqnarray}
&~& \hspace*{-3cm}ds^2 =
\biggl(1-\frac{\mu}{\Sigma\,r^{n-1}}\biggr) dt^2 + \frac{2 a \mu
\sin^2\theta}{\Sigma\,r^{n-1}}\,dt \, d\varphi
-\frac{\Sigma}{\Delta}\,dr^2 -\Sigma\,d\theta^2 \nonumber \\[2mm]
\hspace*{2cm} &-& \biggl(r^2+a^2+\frac{a^2 \mu
\sin^2\theta}{\Sigma\,r^{n-1}}\biggr) \sin^2\theta\,d\varphi^2 -
r^2 \cos^2\theta\, d\Omega_{n}^2, \label{rot-metric}
\end{eqnarray}
where
\begin{equation}
\Delta = r^2 + a^2 -\frac{\mu}{r^{n-1}}\,, \qquad \Sigma=r^2
+a^2\,\cos^2\theta\,, \label{Delta}
\end{equation}
and $d\Omega^2_n$ is the line-element on a unit $n$-sphere. The
mass $M_{BH}$ and angular momentum $J$ of the black hole are then
given in terms of the parameters $a,\mu$
\begin{equation}
M_{BH}=\frac{(n+2) A_{n+2}}{16 \pi G}\,\mu\,,  \qquad
J=\frac{2}{n+2}\,M_{BH}\,a\,, \label{def}
\end{equation}
with $G$ being the $(4+n)$-dimensional Newton's constant, and
$A_{n+2}$ the area of a $(n+2)$-dimensional unit sphere given by
\begin{equation}
A_{n+2}=\frac{2 \pi^{(n+3)/2}}{\Gamma[(n+3)/2]}\,.
\end{equation}
The corresponding line-element on the brane can be found by projecting
out the angular variables that parametrize the extra dimensions. In
that case, the factor $d\Omega^2_n$ disappears and the 4-dimensional
brane background is described by the line-element
\begin{eqnarray}
&~& \hspace*{-5.5cm}ds^2 = \biggl(1-\frac{\mu}{\Sigma\,r^{n-1}}\biggr)
dt^2 +
\frac{2 a \mu \sin^2\theta}{\Sigma\,r^{n-1}}\,dt d\varphi
-\frac{\Sigma}{\Delta}\,dr^2 -\Sigma\,d\theta^2 \nonumber \\[2mm]
\hspace*{4cm}
&-& \biggl(r^2+a^2+\frac{a^2 \mu \sin^2\theta}{\Sigma\,r^{n-1}}\biggr)
\sin^2\theta\,d\varphi^2\,.
\label{rot-metric-4D}
\end{eqnarray}
In the above $0 <\varphi < 2 \pi$, $0< \theta < \pi$ and $n$ stands for
the number of extra, spacelike dimensions that exist transverse to
the brane ($D=4+n$).

We would like to study the emission of gauge bosons and fermions
by the aforementioned projected black-hole background. We assume
that the emitted particle modes couple only minimally to the
gravitational background and have no other interactions,
therefore, they satisfy the corresponding free equations of
motion. By using the Newman-Penrose formalism \cite{NP, Chandrasekhar},
and assuming the factorized ansatz
\begin{equation}
\Psi_s(t,r,\theta,\varphi)= e^{-i\omega t}\,e^{i m \varphi}\,R_{s}(r)
\,S^{m}_{s,j}(\theta)\,,
\label{facto}
\end{equation}
where $S^{m}_{s,j}(\theta)$ are the so-called spin-weighted
spheroidal harmonics \cite{press1, staro, fackerell, breuer, brw,
Casals:2004zq}, the free equations of motion for particles with
spin $s=0,\frac{1}{2}$ and 1 may be combined to form the following
complete ``master" equation \cite{CDKW}, satisfied by the radial
part of all radiative components of the field,
\begin{equation}
\Delta ^{-s}\,
\frac {d}{dr}
\left(\Delta^{s+1}\,
\frac {d R_{s}}{dr}\right)+\left[\frac{K^2-isK\Delta'}{\Delta}+4is\omega r
+s\left( \Delta '' -2 \right)\delta_{s,|s|}-\Lambda_{sj}\right]R_{s}=0\,,
\label{radial}
\end{equation}
where
\begin{equation}
K=(r^2+a^2)\,\omega -a m\,, \qquad \Lambda_{sj}=\lambda_{sj} + a^2
\omega^2
-2 a m \omega\,. \label{lambda}
\end{equation}
In the above, $\lambda_{sj}$ is the angular eigenvalue appearing in the
equation satisfied by the spheroidal harmonics, namely
\begin{equation}
\frac{1}{\sin\theta}\,\frac{d}{d\theta}\left(\sin\theta\,\frac{dS^m_{s,j}}
{d\theta}\right)+ \left(a^2 \omega^{2}\cos^2\theta-2a \omega s\cos\theta
-\frac{(m+s \cos\theta)^{2}}{\sin^2\theta}+ \lambda_{sj}+s\right)
S^m_{s,j}=0\,.
\label{angular}
\end{equation}

From Eq. (\ref{radial}), it is clear that the radial parts of the radiative
components with $s=|s|$ and $s=-|s|$ satisfy different equations due to
the presence of the term that is multiplied by the $\delta_{s,|s|}$
factor. This obstacle can be overcome by redefining
\beq
R_{+|s|} \equiv \Delta^{-|s|}\,P_{+|s|}\,, \qquad
R_{-|s|} \equiv P_{-|s|}\,.
\eeq
In terms of the new radial functions, the master radial equation on
the brane takes the simplified form
\begin{equation}
\Delta^{|s|}\,\frac{d \,}{dr}\,\biggl(\Delta^{1-|s|}\,\frac{d
P_s}{dr}\,\biggr) + \biggl(\frac{K^2-is K\Delta'}{\Delta} +
4 i s \omega r - \tilde \Lambda_{sj}
\biggr)\,P_s(r)=0\,, \label{master2} \eeq
where the $\Delta''$-term has now disappeared, and
$\tilde\Lambda_{sj}=\Lambda_{|s|j}+2 |s|$. The angular eigenvalue
$\lambda_{|s|j}$ cannot be expressed in closed form, however it
can be written in terms of a power series with respect to
$a\omega$ \cite{press1, fackerell, churilov, Seidel}
as\beq\lambda_{|s|j}=-|s|(|s|+1) + \sum_k\, f_k\,(a \omega)^k=
j(j+1)-|s|(|s|+1)-\frac{2ms^2}{j(j+1)}a\omega+...\,.\eeq

By solving Eq. (\ref{master2}) one can compute the absorption
coefficient $|A_{sjm}|^2$ for the propagation of the field on the
specific gravitational background. This quantity is needed to
compute the Hawking radiation of the black hole on the brane. For
example, the differential energy emission rate is given by \beq
\frac {d^{2}E^{(s)}}{dt\,d\omega}  = \frac {1}{2\pi} \sum _{j ,m}
\frac {\omega}{\exp\left[k/T_\text{H}\right] \pm1} |{\cal
A}_{sjm}|^2\, ,\label{eflux} \eeq with $k$ and $T_H$ given by \beq
k \equiv \omega - m \Omega =\omega - \frac{m a}{r_H^2 +a^2}\,,
\qquad \quad
T_\text{H}=\frac{(n+1)+(n-1)a_*^2}{4\pi(1+a_*^2)r_{H}}\,,
\label{k} \eeq with $\Omega$ the angular velocity, $T_H$ the
temperature of the black hole and $r_H$ the black-hole horizon
radius, defined through the relation $\Delta(r_H)=0$.

\section{Analytical Solution}

In this section, we will derive analytic solutions to the radial master
equation valid at the two asymptotic regimes of the black-hole horizon
($ r \simeq r_H$) and far-field ($r \gg r_H$). We will then
demand that the two solutions are smoothly connected at an intermediate
radial zone in order to construct a complete solution, valid at all
radial regimes, for a field with arbitrary spin $s$.

\subsection{The Near-Horizon Regime}

In order to bring Eq. (\ref{master2}) in the form of a known
differential equation, we apply the following transformation \cite{CEKT2}
\beq r \rightarrow f(r) = \frac{\Delta(r)}{r^2+a^2} =
\frac{r^2+a^2-\mu/r^{n-1}}{r^2+a^2}\,\,\Longrightarrow\,
\frac{df}{dr}=(1-f)\,r\,\frac{A(r)}{r^2+a^2}\,, \label{fr}\eeq
where  $A(r)=(n+1)+(n-1)\,a^2/r^2$. Then, we may rewrite Eq.
(\ref{master2}) near the horizon ($r \simeq r_H$) -- keeping $s$
in the master equation as an arbitrary parameter -- as
\begin{equation}
f\,(1-f)\,\frac{d^2P_s}{df^2} + (1-|s|-B_*\,f)\,\frac{d P_s}{df} +
\biggl[\frac{K^2_*-is K_* \Delta_*'}{A_*^2\,f (1-f)} + \frac{(4 i
s \omega_*- \tilde \Lambda_{sj})(1+a_*^2)}{A_*^2\,(1-f)}
\biggr]P_s(r)=0\,, \label{NH-s}
\end{equation}
where we have defined $\omega_*=\omega r_H$, $a_*=a/r_H$. Also
$\Delta_*' = \Delta'(r_H)=A_*$, $B_*$ is now
\beq B_* \equiv 1 -|s| + \frac{2|s|+n\,(1+a_*^2)}{A_*} - \frac{4
a_*^2}{A_*^2}\,, \label{bstar}\eeq
while $A_*$ and $K_*$ are given by
\beq A_*  = n + 1 + (n - 1)a_*^2\,, \label{astar}\eeq
\beq K_*  = (1 + a_* ^2 )\omega _*  - a_* m\,.
\label{kstar}\eeq

If we make the redefinition: $P(f)=f^\alpha (1-f)^\beta F(f)$, Eq.
(\ref{NH-s}) takes the form of a hypergeometric equation \cite{Abramowitz}
\beq
f\,(1-f)\,\frac{d^2 F}{df^2} + [c-(1+a+b)\,f]\,\frac{d F}{df}
-ab\,F=0\,, \label{hyper} \eeq with
\begin{eqnarray}
a=\alpha + \beta +B_*-1\,, \qquad b=\alpha + \beta\,, \qquad c=1
-|s| + 2 \alpha\,.
\end{eqnarray}
The power coefficients $\alpha$ and $\beta$ can be determined by
solving the modified second-order algebraic equations
\beq \alpha^2 -|s|\,\alpha + \frac{K_*^2}{A_*^2}
-\frac{isK_*}{A_*}=0\, \label{alpha-eq-s} \eeq
and
\beq \beta^2 + \beta\,(B_*+|s|-2) + \frac{K_*^2}{A_*^2} -
\frac{isK_*}{A_*} + \frac{(4is\omega_*-
\tilde\Lambda_{sj})\,(1+a_*^2)} {A_*^2}=0\,, \label{beta-eq-s} \eeq
respectively. The first of these two equations leads to the following
solutions for the parameter $\alpha$
\begin{equation}
\alpha_ \pm   = {{|s|} \over 2} \pm \left( {{{iK_* } \over {A_* }} + {s
\over 2}} \right)\,, \label{sol-a}
\end{equation}
while the second equation for $\beta$ admits the solutions
\begin{equation}
\beta_{\pm} =\frac{1}{2}\,\biggl[\,(2-|s|-B_*)\pm \sqrt{(B_*+|s|
-2)^2 - \frac{4 K_*^2-4isK_*A_*}{A_*^2} -  \frac{4(4is\omega_*-
\tilde\Lambda_{sj})\,(1+a_*^2)} {A_*^2}}\,\biggr]\,. \label{beta-s}
\end{equation}
The general solution of the master equation near the horizon is
then given by
\begin{eqnarray}
&& \hspace*{-1cm}P_{NH}(f)=A_- f^{\alpha}\,(1-f)^\beta\,F(a,b,c;f)
\nonumber \\[1mm] && \hspace*{2cm} +\,
A_+\,f^{-\alpha}\,(1-f)^\beta\,F(a-c+1,b-c+1,2-c;f)\,.
\label{NH-gen}
\end{eqnarray}
We must now impose the boundary condition that no outgoing modes
exist near the horizon of the black hole. Using the solutions (\ref{sol-a}),
in the limit $r \rightarrow r_H$, or $f(r) \rightarrow 0$, we obtain either
\beq P_{NH}(f) \simeq A_-\,f^{\frac{|s|+s}{2}}\,f^{iK_*/A_*} +
A_+\,f^{-\frac{|s|+s}{2}}\,f^{-iK_*/A_*}\,,
\end{equation}
for $\alpha=\alpha_+$, or
\beq P_{NH}(f) \simeq A_-f^{\frac{|s|-s}{2}}\,f^{-iK_*/A_*} +
A_+f^{-\frac{|s|-s}{2}}\,f^{iK_*/A_*}\,,
\end{equation}
for $\alpha=\alpha_-$. We may now use the tortoise-like coordinate
$y=r_H (1+a_*^2)\ln(f)/A_*$ that, in the limit $r \rightarrow r_H$, becomes
identical \cite{CEKT2} to the usual tortoise coordinate $r_*$, defined by
$dr_*/dr=(r^2+a^2)/\Delta(r)$. Then, the factors $f^{\pm iK_*/A_*}$
reduce to $e^{\pm i k y}$ describing an outgoing and incoming free wave,
respectively. According to Teukolsky's classic analysis \cite{Teukolsky:1973ha},
the correct boundary condition at the horizon of the black hole, for a
field with non-zero spin, is
\beq R_s \sim \Delta^{-s}\,e^{-i k r_*}\,, \eeq which, in our
case, translates to
\beq
P_{+|s|} \sim e^{-i k y}=f^{-iK_*/A_*}\,, \qquad
P_{-|s|} \sim f^{|s|}\,e^{-i k y}=f^{|s|}\,f^{-iK_*/A_*}\,.
\eeq
Demanding that our asymptotic near-horizon (NH) solution obeys the
above boundary condition, leads to the selection $\alpha=\alpha_-$
and $A_+=0$, bringing the NH solution to the final form
\begin{equation}
P_{NH}(f)=A_-\,f^{\alpha}\,(1-f)^\beta\,F(a,b,c;f)\,.
\label{NH-gen-final}
\end{equation}
The criterion for the convergence of the hypergeometric function
$F(a,b,c;f)$, i.e. $Re(c-a-b)>0$, can finally be applied, and leads
to the choice $\beta=\beta_{-}$. All the above results reduce to
the ones for scalar fields \cite{CEKT2} if we set $s=0$. They also
reduce to the ones valid for general spin-$s$ fields in a non-rotating
black hole background \cite{Kanti2004} if we set $a=0$.

For the purpose of matching the near-horizon and far-field (FF)
solutions at an intermediate radial zone, we need to extrapolate
(`stretch') our NH solution to values of the radial coordinate that are
much larger than the horizon radius. We will do that by changing first the
argument of the hypergeometric function from $f$ to $1-f$ by using
the following relation \cite{Abramowitz}
\bea
P_{NH}(f)&=&A_- f^\alpha\,(1-f)^\beta\,\Biggl[\,\frac{\Gamma(c)\,\Gamma(c-a-b)}
{\Gamma(c-a)\,\Gamma(c-b)}\,F(a,b,a+b-c+1;1-f) \nonumber \\[3mm]
&+& (1-f)^{c-a-b}\,\frac{\Gamma(c)\,\Gamma(a+b-c)}
{\Gamma(a)\,\Gamma(b)}\,F(c-a,c-b,c-a-b+1;1-f)\Biggr].
\eea
The function $f(r)$ may be alternatively written as
\beq
f(r)=1-\frac{\mu}{r^{n-1}}\,\frac{1}{r^2+a^2}=
1-\biggl(\frac{r_H}{r}\biggr)^{n-1}\,\frac{(1+a_*^2)}
{(r/r_H)^2+a_*^2}\,,
\eeq
where we have used the horizon equation  $\Delta(r_H)=0$ in order
to eliminate $\mu$ from the above relation. In the limit $r \gg
r_H$, and for $n \geq 0$, the above expression goes to unity.

By using the above, the argument of the ``stretched"
hypergeometric function goes again to zero, and the ``stretched"
near-horizon solution takes the form
\beq
P_{NH}(f) \simeq A_-\,(1-f)^\beta\,
\frac{\Gamma(c)\,\Gamma(c-a-b)}{\Gamma(c-a)\,\Gamma(c-b)}
+ A_-\,(1-f)^{-\beta+2-B_*-|s|}\,\frac{\Gamma(c)\,\Gamma(a+b-c)}
{\Gamma(a)\,\Gamma(b)}\,. \label{NH-stretched}
\eeq
For $r \gg r_H$, the quantity $(1-f)$ can be accurately approximated by
\beq 1-f \simeq (1+a_*^2)\,\left(\frac{r_H}{r}\right)^{n+1}\,,
\eeq
thus, bringing Eq. (\ref{NH-stretched}) to a simpler power-law form
\beq P_{NH} (r) \simeq A_1\,r^{ - (n + 1)\beta }  +
A_2\,r^{(n + 1)(\beta + |s| + B_*  - 2)}\,, \label{nhstr}\eeq
with
\beq
A_1  = A_ -  \left[ {(1 + a_* ^2 )\,r_H ^{n + 1} } \right]^\beta
{{\Gamma (c)\Gamma (c - a - b)} \over {\Gamma (c - a)\Gamma (c -b)}}\,,
\eeq
\beq A_2  = A_ -  \left[ {(1 + a_* ^2 )\,r_H ^{n + 1} } \right]^{ -
(\beta  + |s| + B_*  - 2)} {{\Gamma (c)\Gamma (a + b - c)} \over
{\Gamma (a)\Gamma (b)}}\,. \label{A1A2}\eeq

\subsection{The Far-Field Regime}

We now turn to the far-field regime, where the radial master equation
(\ref{master2}) becomes
\beq  {{d^2 P_s} \over {dr^2
}} + {{2(1 - |s|)} \over r}{{dP_s} \over {dr}} + \left( {\omega ^2
+ {{2is\omega } \over r} - {{ \lambda _{|s|j}+2|s|+a^2\omega^2 }
\over {r^2 }}} \right)P_s = 0\,. \label{masterff}\eeq
By redefining $P_s= e^{ - i\omega r} r^{{1 \over 2}\left( {2|s| - 1 + Z }
\right)}\tilde P_s$, with
\beq Z=\sqrt{(2|s|-1)^2+4(\lambda_{|s|j}+2|s|+a^2\omega^2)}\,,
\label{zeta}\eeq
Eq. (\ref{masterff}) takes the form of a confluent hypergeometric
differential equation whose solution can be expressed in terms of the
Kummer functions $M$ and $U$, i.e. \cite{Abramowitz}
\bea
&P_{FF}(r) = e^{ - i\omega r} r^{{1 \over 2}(2|s| - 1 + Z )}  \times& \nonumber
\\[3mm]
&  \left[ {B_1\,M\left( {{1 \over 2} - s + {Z \over 2} ,1 +Z
,2i\omega r} \right)} \right.
 + \left. {B_2\,U\left( {{1 \over 2} - s +\frac{Z}{2}  ,1 + Z ,2i\omega r} \right)}\,\right].&
\label{ffsol}\eea As in the case of the NH solution, the FF one
also needs to be ``stretched", this time towards small values of
the radial coordinate. In order to do this, we take the $r
\rightarrow 0$ limit in (\ref{ffsol}), which gives
\cite{Abramowitz}
\beq P_{FF}(r)  \simeq
B_1\,r^{{1 \over 2}(2|s| - 1 + Z )} + B_2\,r^{{1 \over
2}(2|s| - 1 - Z )} {{\Gamma (Z )} \over {\Gamma ({1 \over 2} - s +
\frac{Z}{2}) }}(2i\omega )^{ - Z }\,.\label{FFst}\eeq

In analogy with the stretched NH solution (\ref{nhstr}), the
stretched FF solution also has a power-law form. In order to
construct a complete radial solution, we have to match these two
expressions at an intermediate zone. To this end, we note that,
for $\omega r_H \ll 1$ and $a_* \ll 1$, we get, from Eq.
(\ref{bstar}), $B_* \simeq 2 -|s| + {{2|s|-1} \over {n + 1}} $,
and \beq \beta  \simeq {1 \over 2\,(n+1)}\left( 1-2|s| - \sqrt
{(2|s| - 1)^2 + 4\tilde \Lambda _{sj}}\right). \eeq In this case,
the stretched near-horizon solution (\ref{nhstr}) takes the form
\beq P_{NH} (r) = A_1\,r^{{1 \over 2}\left( {2|s| - 1 + \sqrt
{(2|s| - 1)^2  + 4\tilde \Lambda _{sj} } } \right)}  + A_2\,r^{{1
\over 2}\left( {2|s| - 1 - \sqrt {(2|s| - 1)^2  + 4\tilde \Lambda
_{sj} } } \right)}\,. \label{nhstr2}\eeq From the definitions
(\ref{lambda}) and (\ref{zeta}) for the constants $\Lambda_{|s|j}$
and $Z$, respectively, one can see that, within our approximation,
the powers of $r$ in the two stretched solutions (\ref{FFst}) and
(\ref{nhstr2}) become identical. Then, a smooth matching is realized
provided that we identify the corresponding coefficients in front
of the same power of $r$. From this, we obtain
\beq {{B_1 }
\over {B_2 }} = {{\Gamma (Z )} \over {\Gamma ({1 \over 2} - s
+\frac{Z}{2} ) }}\,(2i\omega )^{ - Z }\,{{A_1 } \over {A_2 }}\,,
\label{B1B2}\eeq
with $A_1/A_2$ following from Eq. (\ref{A1A2})
\beq {{A_1 } \over {A_2 }}
= {{\Gamma (c - a - b)\Gamma (a)\Gamma (b)} \over {\Gamma (c -
a)\Gamma (c - b)\Gamma (a + b - c)}}\left[ {(1 + a_* ^2 )\,r_H ^{n +
1} } \right]^{2\beta + |s| + B_*  - 2}\,. \eeq
Equation (\ref{B1B2}) ensures the smooth matching of the two asymptotic
solutions, and thus the existence of a complete solution to the radial
master equation describing the propagation of a field with arbitrary
spin $s$ in the gravitational background induced on the brane.


\subsection{Computing the Absorption Probability}

The derivation of the complete solution for the radial function $P_s(r)$
opens the way for the calculation of the corresponding absorption
probability $|A_{sjm}|^2$. The latter appears in the expression of the
various emission rates for the Hawking radiation emitted on the brane from
the higher-dimensional black hole given in Eq. (\ref{rot-metric}). In
order to derive this quantity, we expand the far-field solution (\ref{ffsol})
in the limit $r\rightarrow \infty$, and obtain \cite{Abramowitz}
\bea
P_{FF}(r) &\simeq& \left( {{{B_1\,e^{i\pi \left( {{1 \over 2} - s + {Z\over 2} }
\right)} } \over {(2i\omega )^{\left( {{1 \over 2} - s + {Z\over 2} }
\right)} }}{{\Gamma \left( {1 + Z } \right)} \over {\Gamma \left( {{1 \over 2}
+ s + {Z \over 2} } \right)}} + } \right.
\left. {  {{B_2 } \over {(2i\omega )^{  \left( {{1 \over 2} - s +
{Z \over 2} } \right)} }}} \right) \frac{e^{ - i\omega r}}{r^{1-s-|s|}}
\nonumber \\[2mm]
&+& {{B_1 } \over {(2i\omega )^{\left( {{1 \over 2} + s + {Z \over 2} }
\right)} }}{{\Gamma \left( {1 + Z } \right)} \over {\Gamma \left( {{1 \over 2}
- s + {Z\over 2} } \right)}}\,\frac{e^{i\omega r}}{r^{1+s-|s|}}
\nonumber \\[2mm]
&\equiv& Y_s ^{(in)} {{e^{ - i\omega r} } \over {r^{1 - s - |s|} }} +
Y_s ^{(out)} {{e^{i\omega r} } \over {r^{1 + s - |s|} }}\,. \label{ffinf}
\eea

Let us focus first on the case of fields with spin $1/2$: from the master
radial equation (\ref{master2}) one may easily derive, similarly to the
4-dimensional case \cite{Chandrasekhar, Unruh, Guven, Leahy}, that
\beq
\frac{d}{dr}\left(|P_{\frac{1}{2}}|^2-|P_{-\frac{1}{2}}|^2\right)=0\,.
\eeq
The conserved -- for arbitrary values of $r$ -- quantity inside
the parenthesis is proportional to the radial component of the
particle current produced by the black hole. Then, the absorption
probability is defined as the ratio of the flux of particles at
the black hole horizon over the one at infinity
\beq
{|A}_{{\frac{1}{2} jm}}|^2 = \frac{F_{in}^{(H)}}{F_{in}^{(\infty)}}=
1-\frac{F_{out}^{(\infty)}}{F_{in}^{(\infty)}}\,,
\eeq
where in the second part of the equation, we have used the conservation
of the total flux. The flux of fermions at infinity may be found by
integrating the radial component of the conserved current over a
2-dimensional sphere at asymptotic infinity. Applying this and using
Eq. (\ref{ffinf}), we find
\beq
|A_{\frac{1}{2} jm}|^2 = 1-\frac{|P_{\frac{1}{2}}^{(out)}|^2-
|P_{-\frac{1}{2}}^{(out)}|^2} {|P_\frac{1}{2}^{(in)}|^2-
|P_{-\frac{1}{2}}^{(in)}|^2} =
1+\frac{|Y_{-\frac{1}{2}}^{out}|^2}{|Y_\frac{1}{2}^{(in)}|^2}\,.
\eeq
Using, finally, the explicit expressions for $Y_{\pm \frac{1}{2}}^{(out)}$, as
these are defined in Eq. (\ref{ffinf}), we find that they are related
through the equation
\beq
Y_{-\frac{1}{2}}^{(out)} =
{{2i\omega } \over {\sqrt {\lambda _{1/2j}+1+a^2\omega^2  }
}}Y_\frac{1}{2} ^{(out)}\,, \label{Y1/2}
\eeq
which, therefore, leads to the following expression for the absorption
probability for fermions
\beq
|A_{\frac{1}{2} jm}|^2 = 1 - {{4\omega ^2 } \over {\lambda
_{1/2j}+1+a^2\omega^2 }}\left| {{{Y_{\frac{1}{2}} ^{(out)} } \over
{Y_{\frac{1}{2}} ^{(in)} }}} \right|^2\,. \label{A12}\eeq

For fields with spin $s=1$, there is no conserved particle current.
In order to compute the absorption probability, a technique, introduced
in \cite{Detweiler}, may be followed in which the radial master equation
is transformed, through a radial function redefinition and the use of
the tortoise coordinate $r_*$, to an alternative one with real, short-range
potential. Then, the asymptotic solution at infinity for the gauge field
is given, in terms of the new radial function, by the following expression
\cite{Detweiler, CO}
\beq
X_{jm\omega} \sim e^{-i \omega r_*} + A^{(in)}_{jm\omega}\,e^{i \omega r_*}\,,
\label{det}
\eeq
i.e. by the sum of outgoing and incoming plane waves with constant
amplitudes, from which the expression of $A_{1jm}$ may easily follow
\beq
|A_{1jm}|^2=1-|A^{(in)}_{jm\omega}|^2\,.
\eeq
Although the exact analysis is quite cumbersome \cite{Detweiler,
CO}, it yields relations connecting the constant amplitude
$A^{(in)}_{jm\omega}$ in Eq. (\ref{det}) with the constant
coefficients $Y_s^{(in,out)}$ appearing in our Eq. (\ref{ffinf})
and leading finally to
\beq
|A_{1jm}|^2 = 1 - \frac{16\omega ^4 }{B^2_{jm\omega}}\left|\frac{Y_1 ^{(out)}}
{Y_1^{(in)}}\right|^2. \label{A1}\eeq
The constant $B_{jm\omega}$ is defined as the coefficient appearing in the
differential equation \cite{Chandrasekhar}
\beq
\Delta \, D_0^\dagger D_0^\dagger P_{+1}=B_{jm\omega}\,P_{-1}\,,
\label{Bjmw}
\eeq
where $D_0^\dagger=\partial_r + i K/\Delta$, or, equivalently,
through the relation
\beq
Y_{ - 1} ^{(out)} =  - \frac{4\omega ^2 }{B_{jm\omega}}\,
Y_1^{(out)} \label{Y}
\eeq
holding for the asymptotic solution (\ref{ffinf}) when substituted
in Eq. (\ref{Bjmw}). By using the explicit expressions of $Y_{\pm 1}^{(out)}$
from Eq. (\ref{ffinf}), we obtain
\beq
Y_{ - 1} ^{(out)} =  - \frac{4\omega ^2 }{\lambda _{1j}+2+a^2\omega^2 }\,
Y_1^{(out)}\,, \label{Y1}
\eeq
which leads to $B_{jm\omega}=\lambda_{1j}+2+a^2\omega^2$, and,
thus, to the final expression for the absorption probability for
brane-localised fields with spin 1
\beq
|A_{1jm}|^2  = 1 - \frac{16\omega ^4}{(\lambda_{1j}+2+a^2\omega^2)^2}
\left|\frac{Y_1^{(out)}}{Y_1^{(in)}}\right|^2. \label{A1-final}\eeq


\section{Plotting our Analytic Results}

\begin{figure}[t]
  \begin{center}
  \mbox{\includegraphics[width = 0.5 \textwidth] {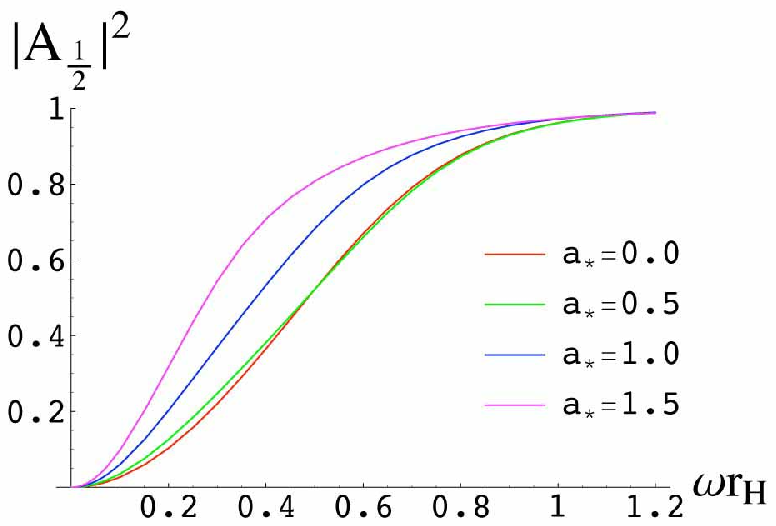}}
\hspace*{-0.3cm} {\includegraphics[width = 0.5 \textwidth]
{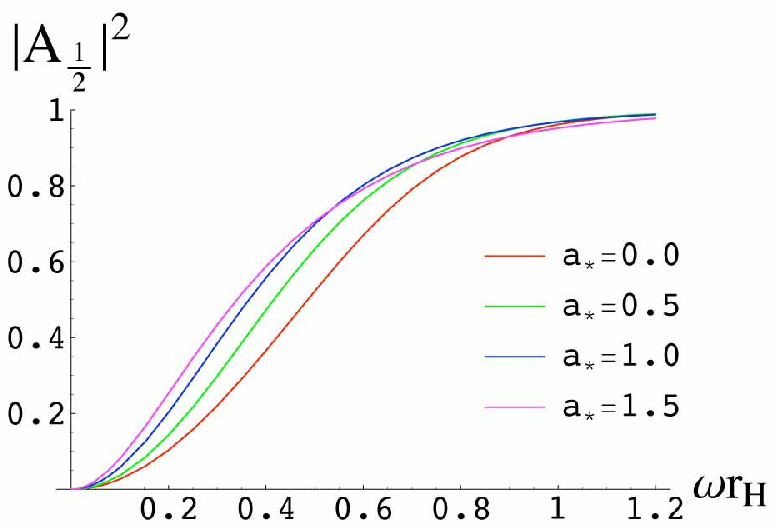}} \caption{ Absorption probability $|{\cal
A}_{1/2}|^2$ for brane spinor particles, for the modes $j=1/2$ and
$m=1/2,-1/2$, from left to right, for $n=6$ and $a_*=$ 0.0, 0.5,
1.0, 1.5.}
    \label{fig-spA}
  \end{center}
\end{figure}
\begin{figure}[ht]
  \begin{center}
  \mbox{\includegraphics[width = 0.33 \textwidth] {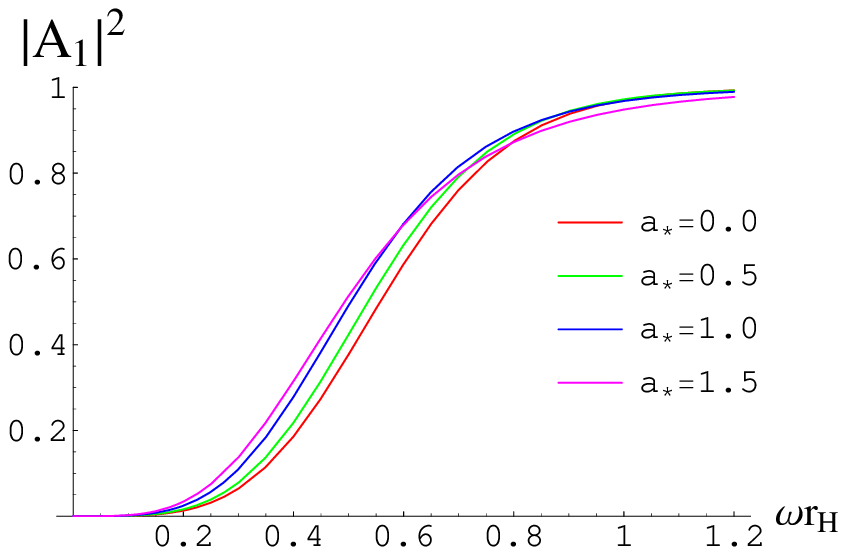}}
\hspace*{-0.3cm} {\includegraphics[width = 0.33 \textwidth]
{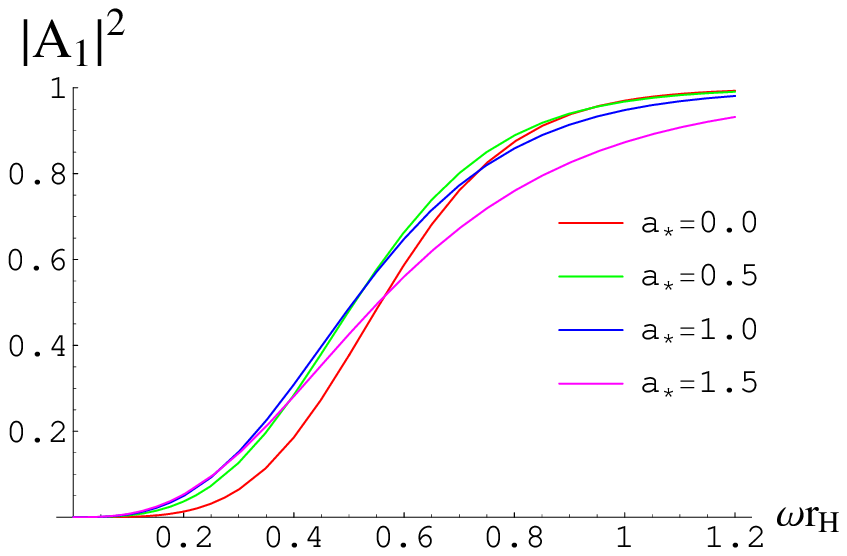}\includegraphics[width = 0.33 \textwidth]
{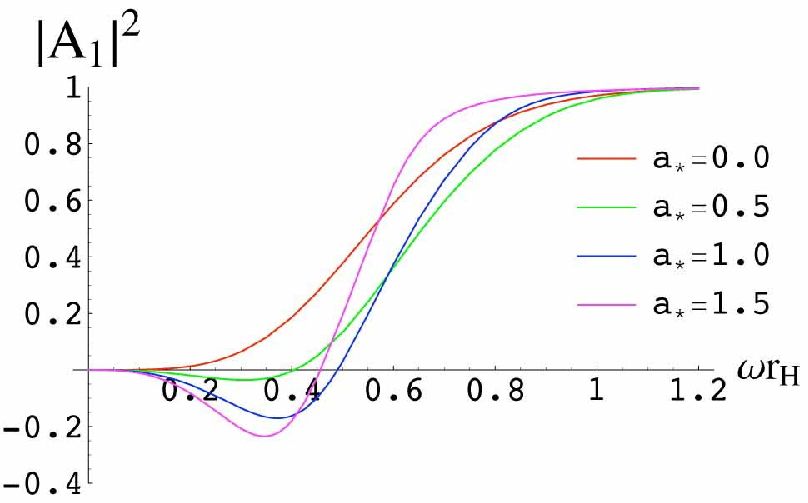}}
    \caption{ Absorption probability $|{\cal A}_{1}|^2$ for brane boson
    particles, for the modes $j=1$ and $m=0,-1,1$, from left to right, for $n=6$ and
    $a_*=$ 0.0, 0.5, 1.0, 1.5.}
    \label{fig-vecA}
  \end{center}
\end{figure}

We can now use Eqs. (\ref{A12}) and (\ref{A1-final}) -- our main
analytic results -- to produce plots for the absorption
probabilities for spin-1/2 and spin-1 particles, respectively,
propagating in the induced-on-the-brane gravitational background.
Upon selection
of particular values of the angular momentum numbers ($j,m$), the
values of $A_{sjm}$ will be plotted as a function of the energy
parameter $\omega r_H$, and then compared with existing numerical
results from the literature \cite{CKW, CDKW, IOP2}.

We remind the reader that in the process of producing a complete
analytic solution to the radial master equation, we were forced to
assume that $\omega r_H \ll 1$ and $a_* \ll 1$, therefore, strictly
speaking, our results are valid only in the low-energy and low-rotation
limit. However, at times, our plots will extend beyond this range of
validity to exhibit the remarkably good agreement we obtain even outside
the assumed range of validity of our approximations. Another point that we
would like to stress is that, in general, an increase in the number of
extra dimensions $n$ improves the validity of our approximation: as $n$
increases, the assumed behavior of $f(r)$ at infinity in Eq. (\ref{fr})
becomes more accurate, and terms that were neglected during the matching
of the asymptotic solutions, such as $K_*/A_*$, become even more suppressed.

In Figs. \ref{fig-spA} and \ref{fig-vecA}, we plot the absorption
probability for fermions and gauge bosons, respectively, for the lowest
partial waves in each case, i.e. ($j=1/2, m=\pm 1/2$) and ($j=1, m=0,-1,1$),
in the case of $n=6$ and for various values of $a_*$. From Fig. \ref{fig-spA},
we see that, for fermions with $m>0$, the absorption probability is
monotonically increasing with the angular momentum parameter over most of
the energy regime, while, for $m<0$, $|A_{1/2}|^2$ increases with $a_*$
in the low-energy regime and is suppressed at the high-energy regime.
For bosons, and $m=0$, the absorption probability is increasing with
the angular momentum of the black hole for small values of $a_*$ but
this increase is decelerated (and eventually reversed) for higher values
of $a_*$; for $m<0$, the absorption probability is enhanced with $a_*$
in the low-energy regime but suppressed in the high-energy one; finally
for $m>0$, an increase in $a_*$ leads to more negative values of
$|A_1|^2$ in the superradiant regime \cite{super} (where the quantity
$\omega- m \Omega$, as well as $|A_1|^2$, becomes negative signalling
the enhancement of the incoming wave's amplitude by a rotating black hole)
and to more positive values in the non-superradiant one.
Our results are in excellent agreement with the exact numerical ones
\cite{CKW, CDKW, IOP2} in the low-energy and low-angular-momentum regime,
and remain in very good agreement even beyond this range of values
-- a deviation from the exact results appears only for values of the
angular momentum parameter larger than unity.

A similar agreement with the exact behaviour appears in the dependence
of the absorption probabilities for both fermions and gauge bosons on the
number of extra dimensions $n$. To avoid repetition of known results,
we do not present here any plots for the dependence of $|A_{1/2}|^2$ and
$|A_1|^2$ on $n$, but only comment on the derived behaviour. For fermions, a
monotonic decrease is observed for $m<0$, while for $m>0$, an enhancement
in the low-energy regime is followed by a suppression in the high-energy
one. For bosons, a monotonic suppression with $n$ is found for the
non-superradiant modes ($m\leq 0$) over the whole energy regime, while
for the superradiant ones ($m>0$) a similar suppression in the non-superradiant
regime is replaced by an enhancement (i.e. more negative values) in
the superradiant one.


\section{Asymptotic Values and Absorption Cross-Section}

By expanding further our analytic expressions (\ref{A12}) and
(\ref{A1-final}) for the absorption probabilities for fermions
and gauge fields, respectively, one may obtain simplified, compact
expressions that reveal more clearly the low-energy asymptotic
behaviour in each case as well as potential differences in the
asymptotic values of the corresponding absorption cross-sections.

\begin{figure}[t]
  \begin{center}
  \mbox{\includegraphics[width = 0.5 \textwidth] {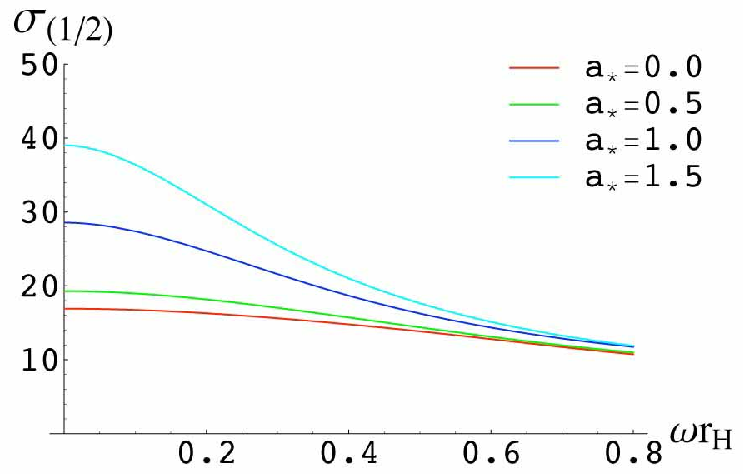}}
\hspace*{-0.4cm} {\includegraphics[width = 0.5 \textwidth]
{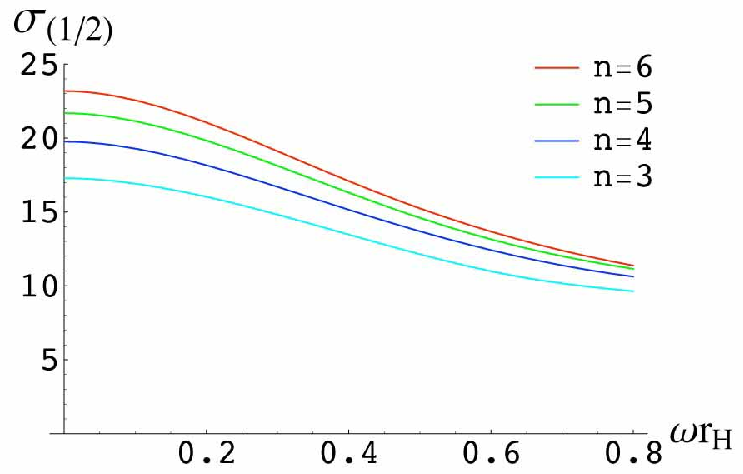}}
    \caption{Total absorption cross-section for spinor fields, in $r_H^2$ units,
    for the cases (a) $n=6$ and various $a_*$, and (b) $a_*=0.5$ and
    various $n$, in low energies.}
    \label{cslow-ferm}
  \end{center}
\end{figure}

In the limit $\omega \rightarrow 0$, we obtain, from Eq. (\ref{zeta}),
$Z \simeq 2j+1$, and then, from the expression of $B_1/B_2$,
Eq. (\ref{B1B2}), the result
\beq
\frac{B_1}{B_2}  \simeq {{\Gamma (2j + 1)} \over {\Gamma (1 + j - s)}}\,
(2i\omega )^{- 2j - 1}\,\left(\frac{A_1}{A_2}\right)_{\omega=0}
\equiv M_{sjm}\,(2 i\omega) ^{ - 2j - 1}\,,
\label{b0lim} \eeq
with $M_{sjm}$ a complex constant independent of the energy $\omega$. Starting from the fermions and Eq. (\ref{A12}), we
set $s=1/2$ and find
\bea
|A_{\frac{1}{2}jm}|^2 &\simeq& 1 - \left|\frac{M_{\frac{1}{2}jm}\,\Gamma (2j + 2)}
{M_{\frac{1}{2}jm} \Gamma (2j + 2)\,e^{i\pi (j + 1/2)} +
\Gamma (j + 3/2)\,(2 i\omega)^{2j + 1}}\right|^2 \nonumber \\[2mm]
& \simeq& \frac{(2 \omega)^{2 j+1}\,\Gamma(j+3/2)}{\Gamma(2 j+2)}
\left(\frac{1}{M_{\frac{1}{2}jm}} + \frac{1}{M^*_{\frac{1}{2}jm}}\right)
+ .... \,.
\eea
The corresponding absorption cross-section, defined as
$\sigma_{sjm}=\pi\,|A_{sjm}|^2/\omega^2$ for each par\-tial wave \cite{cross-section},
will then have the form
\beq
\sigma_{\frac{1}{2}jm}=\frac{2^{2j+1}\pi\,\omega^{2 j-1}\,\Gamma(j+3/2)}{\Gamma(2 j+2)}
\left(\frac{1}{M_{\frac{1}{2}jm}} + \frac{1}{M^*_{\frac{1}{2}jm}}\right) + ...\,.
\label{cross-ferm}
\eeq
From the above result, we may easily read that, similarly to the case
of a non-rotating higher-dimensional black hole \cite{kmr1, HK1}, the
absorption cross-section of the lowest fermionic mode, with $j=1/2$,
assumes a non-zero asymptotic value, namely
\beq
\sigma_{\frac{1}{2}\frac{1}{2}m}=2 \pi \left(\frac{1}{M_{\frac{1}{2}
\frac{1}{2}m}} + \frac{1}{M^*_{\frac{1}{2}\frac{1}{2}m}}\right)\,,
\eeq
while all higher fermionic partial waves, with $j > 1/2$, will have
zero absorption cross-section as $\omega \rightarrow 0$. The quantity
$M_{\frac{1}{2}\frac{1}{2}m}$ depends both on the number of extra dimensions
$n$ and on the angular momentum parameter $a_*$ of the black hole. In
Figs. \ref{cslow-ferm}(a,b), we depict the dependence of
$\sigma_{\frac{1}{2}}$ -- summed over $j$ and $m$ -- on both $a_*$
and $n$, respectively. We observe that the asymptotic value of the
absorption cross-section for fermions in the low-energy regime is
enhanced in terms of both parameters of the gravitational background.

\begin{figure}[t]
  \begin{center}
  \mbox{\includegraphics[width = 0.5 \textwidth] {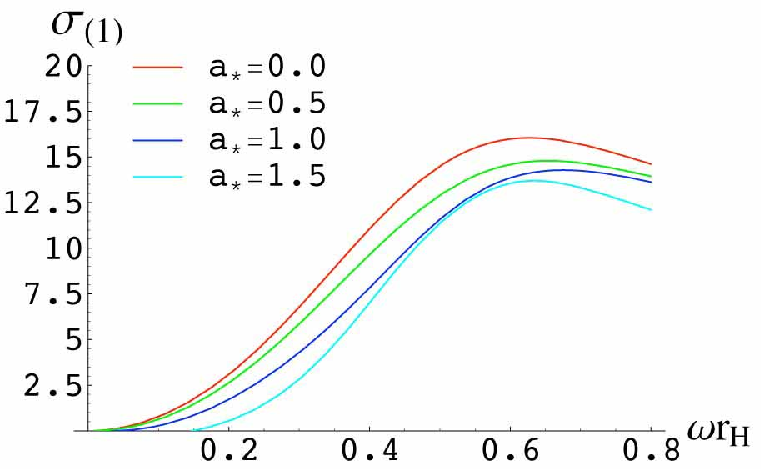}}
\hspace*{-0.3cm} {\includegraphics[width = 0.5 \textwidth]
{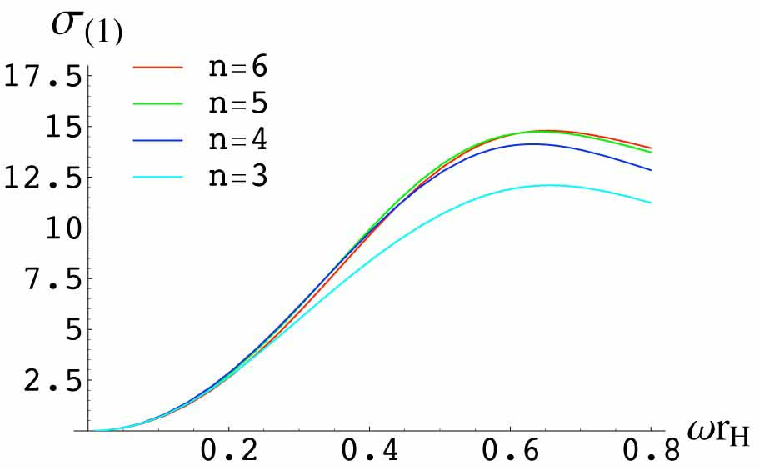}}
    \caption{Total absorption cross-section for vector fields, in $r_H^2$ units,
    for the cases (a) $n=6$ and various $a_*$, and (b) $a_*=0.5$ and
    various $n$, in low energies.}
    \label{cslow-gb}
  \end{center}
\end{figure}

For gauge fields, with $s=1$, Eq. (\ref{A1-final}) leads to a similar
result for the absorption probability, namely
\bea
|A_{1jm}|^2 &\simeq& 1 - \left|\frac{M_{1jm}\,\Gamma (2j + 2)}
{M_{1jm} \Gamma (2j + 2)\,e^{i\pi j} +
\Gamma (j + 2)\,(2 i\omega)^{2j + 1}}\right|^2 \nonumber \\[3mm]
& \simeq& \frac{(2 \omega)^{2 j+1}\,\Gamma(j+2)}{\Gamma(2 j+2)}
\,\frac{i(M^*_{1jm}-M_{1jm})}{|M_{1jm}|^2} + .... \,.
\eea
The corresponding absorption cross-section will similarly have the form
\beq
\sigma_{1jm}=\frac{2^{2j+1}\pi\,\omega^{2 j-1}\,\Gamma(j+2)}{\Gamma(2 j+2)}
\,\frac{i(M^*_{1jm}-M_{1jm})}{|M_{1jm}|^2} + ...\,.
\label{cross-gb}
\eeq
In this case, all gauge fields modes, including the lowest one with
$j=1$, have zero asymptotic value of absorption cross-section -- this
is in agreement with the corresponding results derived in the non-rotating
case \cite{kmr1, HK1}. The dependence of $\sigma_{1}$ -- summed again over
$j$ and $m$ -- on $a_*$ and $n$ is now given in Figs. \ref{cslow-gb}(a,b).
Again, an increase in both parameters causes an enhancement in the absorption
cross-section for gauge fields in the low-energy regime.

As we saw in the previous section, the absorption probability $|A_{sjm}|^2$
for both species of fields, fermions and gauge bosons, approaches unity for
large values of the energy parameter $\omega r_H$. Therefore, the
absorption cross-section for each partial wave will start from either
a zero or non-zero low-energy value -- depending on the spin of the
field and its partial wave numbers, will reach a maximum value and then
asymptote to zero as ${\cal O}(1/\omega^2)$. Nevertheless, we expect the
behaviour of the total cross-section $\sigma_s$ -- summed over $j$ and
$m$ -- to be radically different. Analyses in the case of a non-rotating
black hole \cite{HK1, Emparan} have demonstrated that the total cross-section
for all brane-localised species of fields -- scalar, spinor and gauge
fields -- reach asymptotically a universal constant high-energy
value that depends only on the number of extra dimensions. In the
case of a rotating black hole, it has been similarly shown, both
in the 5-dimensional \cite{Jung-rot} and $(4+n)$-dimensional case
\cite{CEKT2} that, in the case of scalar fields, a similar constant
asymptotic value is reached in the high-energy regime. In the latter
work \cite{CEKT2}, it was shown in detail that the absorption cross-section
can be described in the high-energy regime by the geometrical optics limit
and that the exact asymptotic value of $\sigma_s$ can be approximated by an
analytic expression giving the cross-section for a particle approaching the
black hole parallel to the rotation axis.

\begin{figure}[t]
  \begin{center}
  \mbox{\includegraphics[width = 0.5 \textwidth] {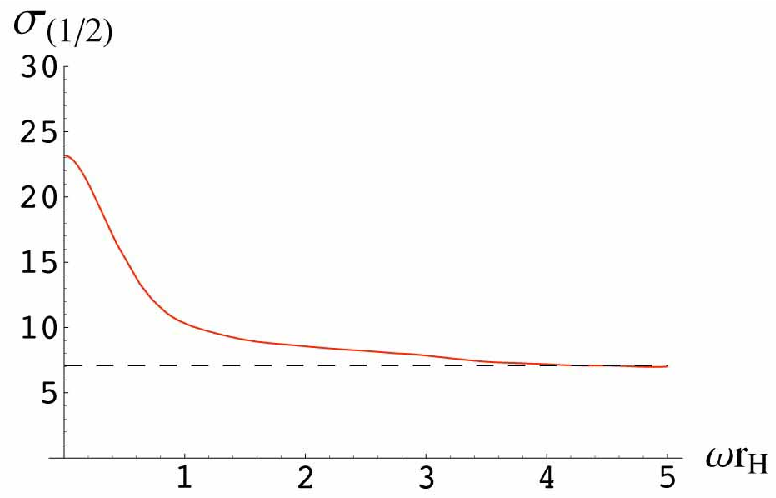}}
\hspace*{-0.3cm} {\includegraphics[width = 0.5 \textwidth]
{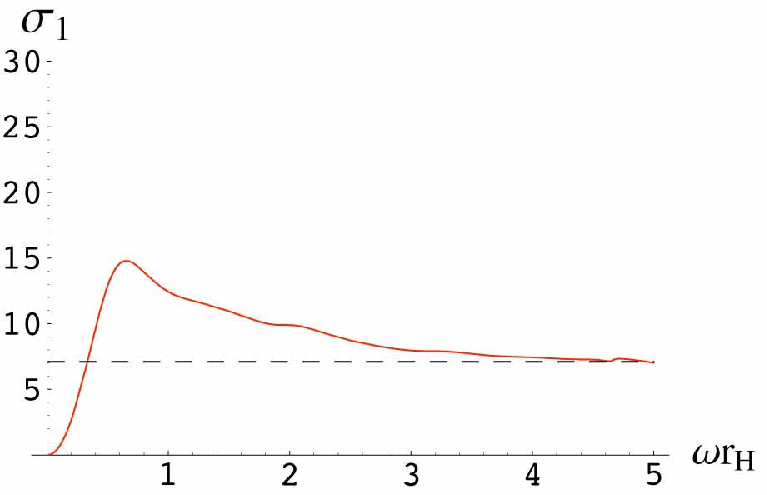}}
    \caption{Total cross section, in $r_H^2$ units, for spinor and vector
    fields, for the case $n=6$ and $a_*=0.5$. The dashed line denotes the
    theoretical asymptotic value of $\sigma$, as computed in \cite{CEKT2}. }
    \label{geomlimit}
  \end{center}
\end{figure}

Thus, it remains to be seen whether the absorption cross-section, in the
case of the remaining species of fields -- i.e. fermions and gauge bosons,
approaches again the same asymptotic value at high energies, as in the
non-rotating case. By using our analytic expressions (\ref{A12}, \ref{A1-final})
for the absorption probabilities, we have computed the partial wave cross-sections
and, by taking the sum over $j$ and $m$ in each case, the total absorption
cross-sections for fermions and gauge fields, respectively. Their behaviour
over the whole energy regime is shown in Figs. \ref{geomlimit}(a,b).
Although our results are expected to hold only in the
low-energy limit and therefore not to yield any reliable results in
the high-energy regime, we have nevertheless attempted to compute the
high-energy asymptotic value of $\sigma_s$. Surprisingly enough, our
results do not break down at high $\omega r_H$ and successfully yield
a constant high-energy asymptotic value for the absorption cross-sections
$\sigma_{1/2}$ and $\sigma_1$. In addition, the asymptotic values are
identical to the one for scalar fields computed analytically in \cite{CEKT2}
-- and denoted here with the horizontal dashed line. Our results, therefore,
prove the universality in the behaviour of fields with arbitrary
spin in the high-energy regime, and the dependence of their absorption
cross-section only on parameters of the gravitational background even
in the case of a higher-dimensional rotating black hole.

\section{Energy Emission Rates}

As is well known, the absorption probabilities $|A_{sjm}|^2$ can be
used in order to determine the various emission rates for the Hawking
radiation. In the present case, our derived analytic expressions
(\ref{A12}) and (\ref{A1-final}), together with the expression for
the temperature (\ref{k}), can lead to the emission rates for
a higher-dimensional rotating black hole directly on our brane, in the
form of fermions and gauge bosons.

\begin{figure}[t]
  \begin{center}
  \mbox{\includegraphics[width = 0.5 \textwidth] {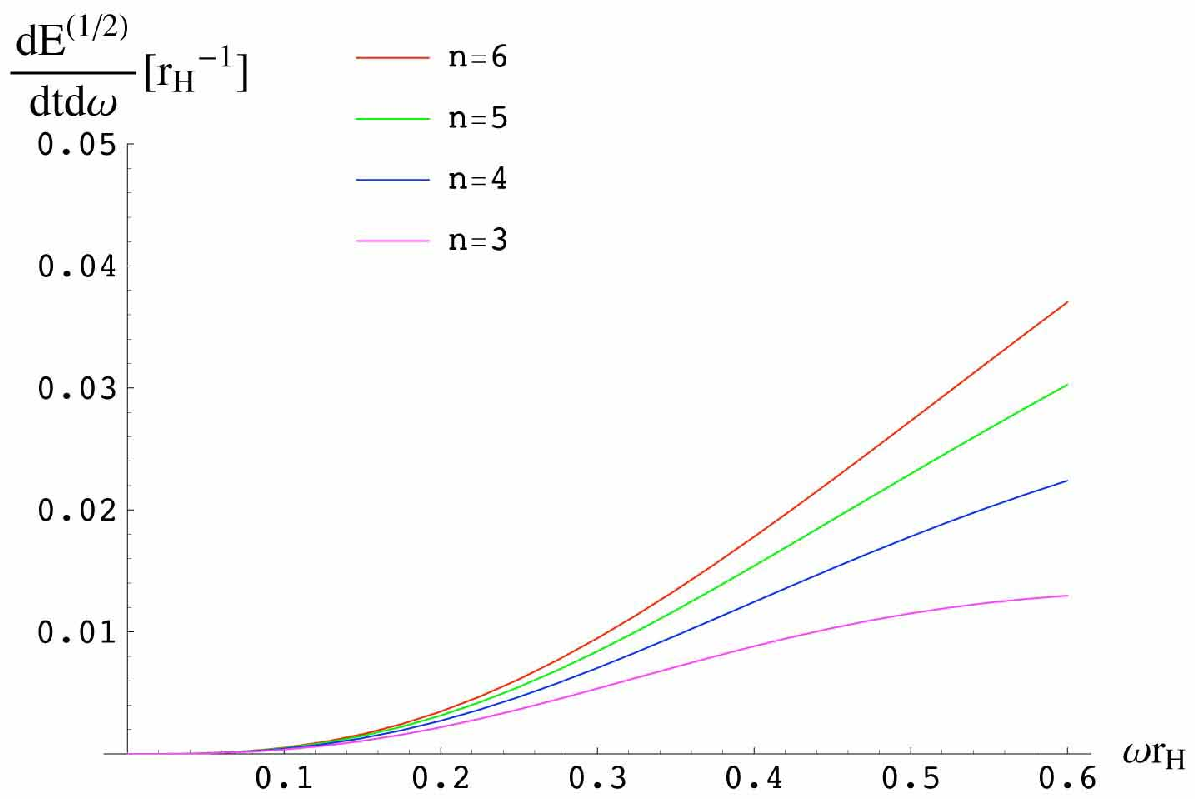}}
\hspace*{-0.3cm} {\includegraphics[width = 0.5 \textwidth]
{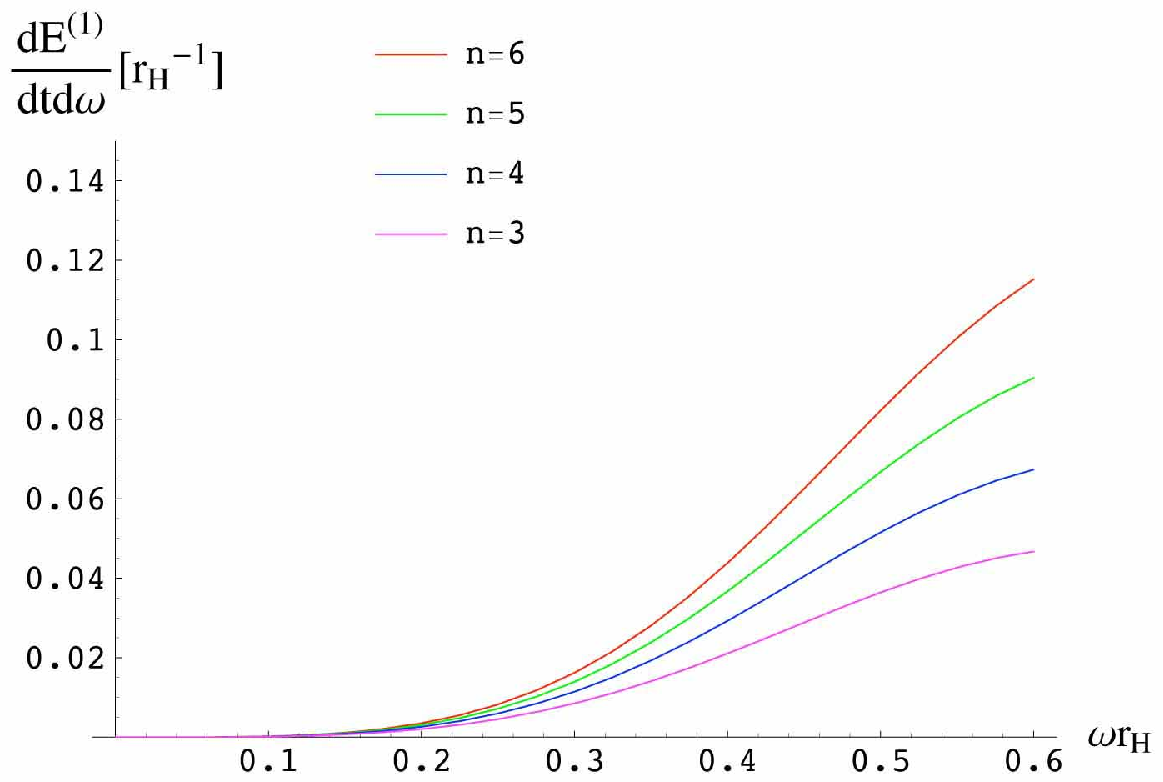}}
    \caption{ Energy emission rates for spinor and vector fields, for $a_*=0.5$
    and various $n$.}
    \label{fig-en-a05}
  \end{center}
\end{figure}
\begin{figure}[ht]
  \begin{center}
  \mbox{\includegraphics[width = 0.5 \textwidth] {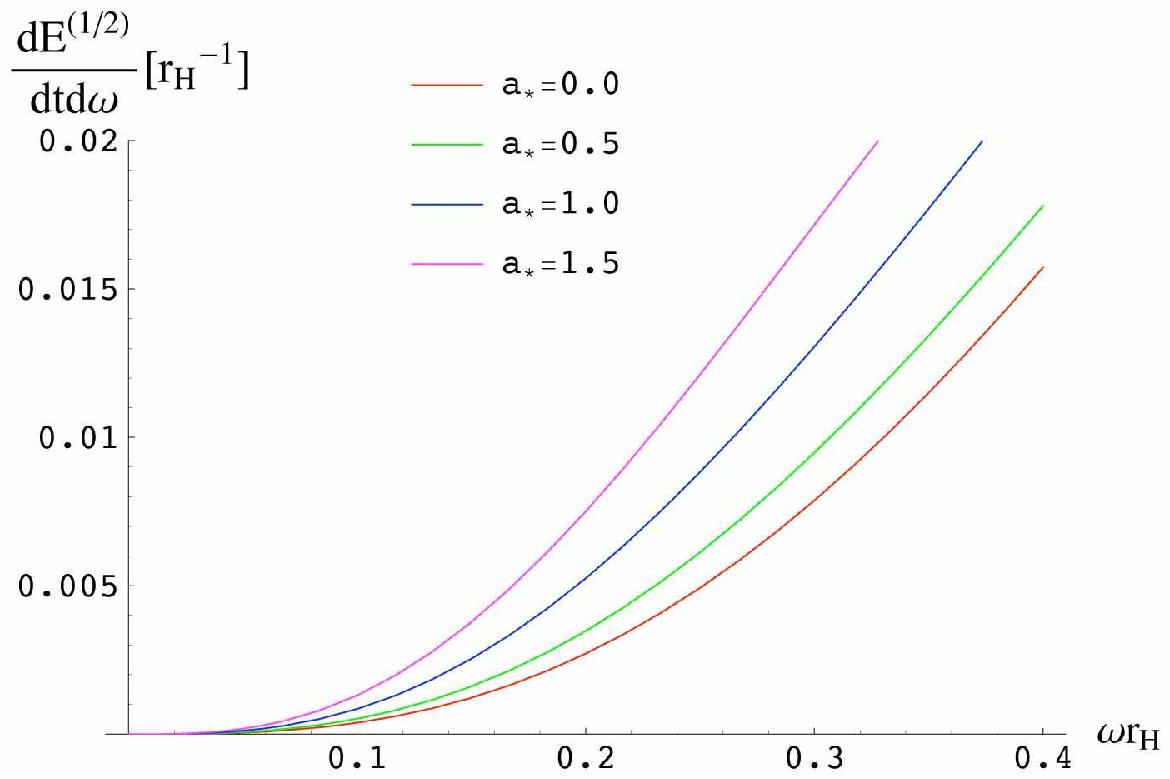}}
\hspace*{-0.3cm} {\includegraphics[width = 0.5 \textwidth]
{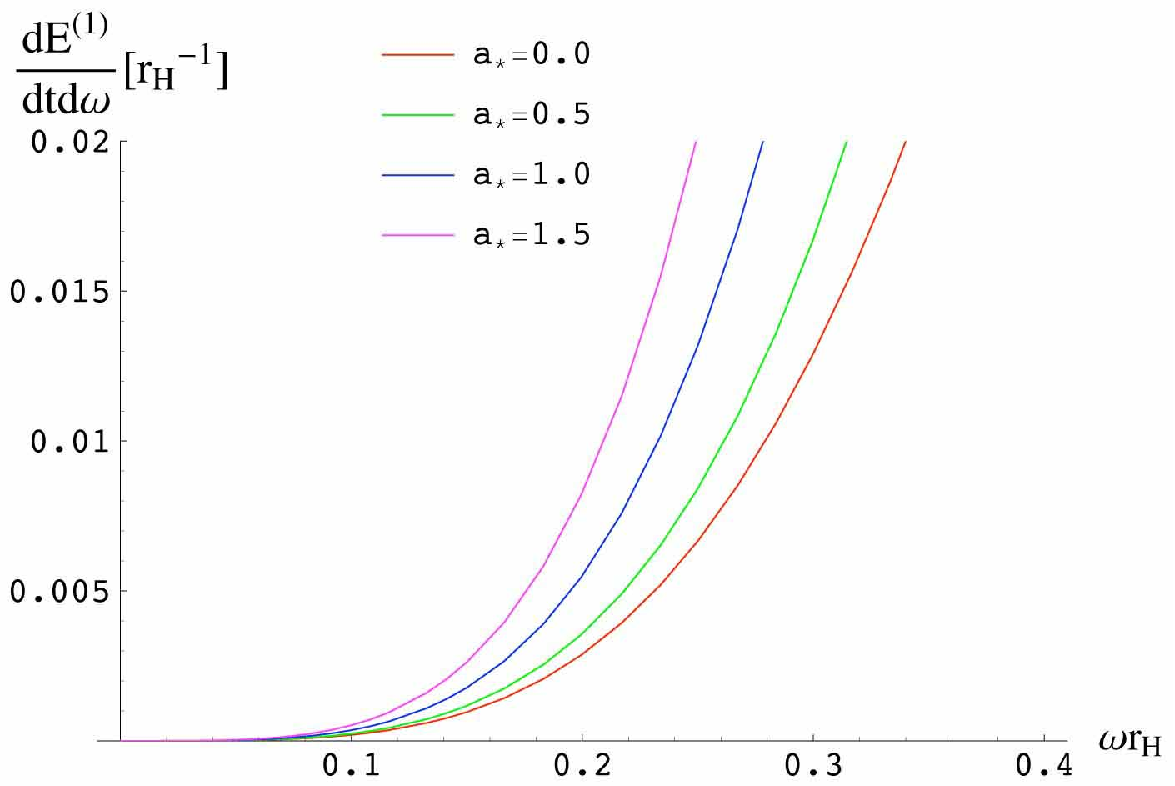}}
    \caption{ Energy emission rates for spinor and vector fields, for $n=6$ and
    various $a_*$.}
    \label{fig-en-n6}
  \end{center}
\end{figure}

Since our analytic expressions are, strictly speaking, valid in
the low-energy and low-angular-momentum limit, we first focus our
attention on these particular regimes. In the context of our
analysis, we have studied all emission rates, associated with the
Hawking radiation, namely the flux (number of particles), power
(energy) and angular momentum emission rates, and they were all
found to exhibit the same behaviour in terms of the number of
extra dimensions $n$ and the angular momentum parameter $a_*$.
Therefore, as an indicative example, we display here our results
for the energy emission rates following from Eq. (\ref{eflux}). In
Figs. \ref{fig-en-a05} and \ref{fig-en-n6}, we present the energy
emission rate, i.e. the energy emitted by the black hole per unit
time and unit frequency, in the form of fermions and gauge fields,
as a function of $n$ and $a_*$, respectively. One may easily
observe that the energy flux is indeed enhanced with both
topological parameters, in accordance with the exact numerical
results of \cite{CKW, CDKW, IOP2}.

\begin{figure}[t]
  \begin{center}
  \mbox{\includegraphics[width = 0.5 \textwidth] {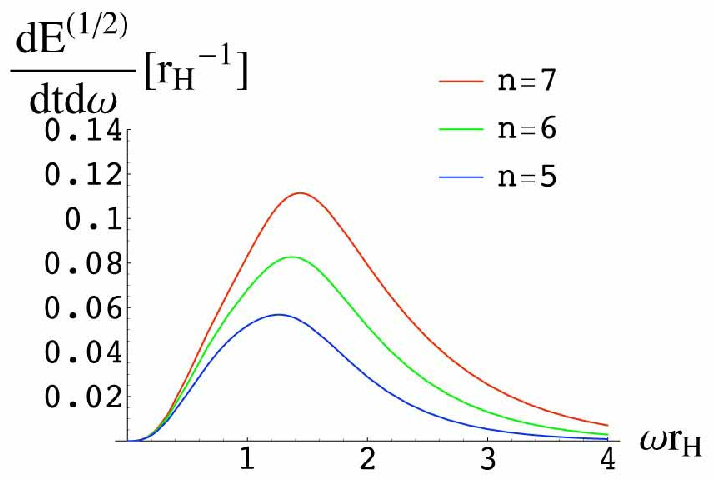}}
\hspace*{-0.3cm} {\includegraphics[width = 0.5 \textwidth]
{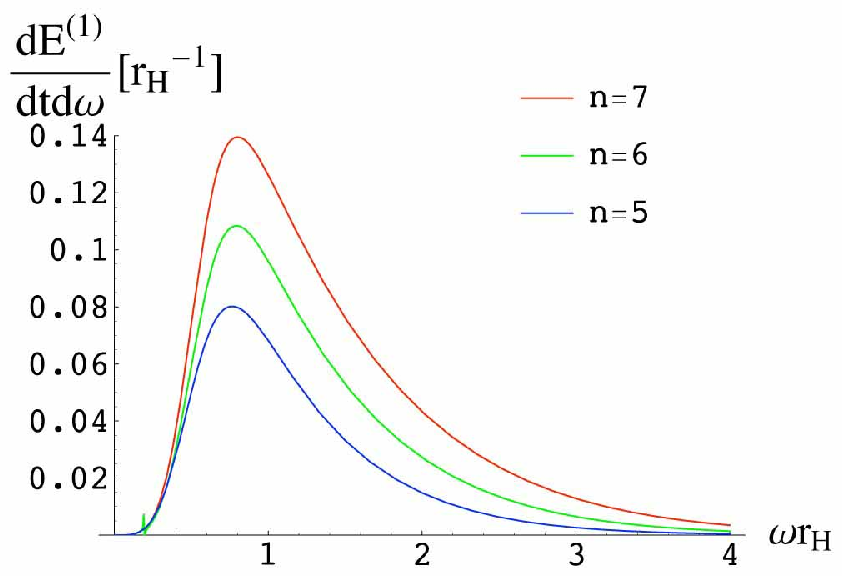}}
    \caption{ Energy emission rates for spinor and vector fields, for $a_*=0.2$
    and various $n$.}
   \label{fig-allen-a02}
  \end{center}
\end{figure}

Even by looking at the low-energy regime, one can accurately conclude
that the emission of gauge bosons dominates over the one for fermions,
for the same values of the parameters $a_*$ and $n$. This feature is
also in accordance with the results of the exact numerical analysis.
In addition, the magnitudes of the corresponding energy emission rates
are also correctly reproduced by our analytic results. Motivated by
this, in Fig. \ref{fig-allen-a02}, we plot the energy emission rates,
for fermions and gauge bosons, for higher values of the energy parameter
$\omega r_H$ and various configurations of $n$, while $a_*$ is kept fixed.
The restricted validity of our analytic results are bound to lead to
deviations from the exact numerical results: indeed, our curves tend
to reach their maximum point and their tail region sooner than the
exact ones. Nevertheless, apart from the above, the agreement is
remarkable: not only is the general profile of the energy emission
rates and their dependence on the number of extra dimensions reproduced
but the magnitudes at the peak of their curves -- reached at
$\omega r_H \simeq 1$, or even well beyond this -- are also accurately
derived.

\section{Conclusions}

In the present article, we studied the emission of Hawking radiation by
higher-dimensional rotating black holes into fermionic and gauge boson
degrees of freedom on the brane. The same topic has been studied before
in the literature, however, those studies were either purely numerical
or focused on specific cases like 5-dimensional spacetimes. In contrast,
here, we carried out an analytical study of the problem deriving results
valid for a black hole in a spacetime with arbitrary numbers of
extra dimensions.

As a starting point of our analysis, we formulated the radial master
equation, describing the propagation of degrees of freedom with spin $s$
on the brane, in such a way that both radiative components of the given
fields, i.e. with $s=\pm|s|$, are described through the same equation.
As the complexity of the problem forbids an all-energies analytical treatment,
we followed an approximate method that allows us to find a solution
in the low-energy and low-rotation regime. In Section 3, we used a well-known
matching technique, which consists of solving the field equations in the
near-horizon and far-field regime, and then matching the two solutions
to construct a complete, smooth solution for the radial part of the field.
This allowed us to compute the absorption probabilities for brane-localised
fermions and gauge bosons through two purely analytic expressions,
namely Eqs. (\ref{A12}) and (\ref{A1-final}), respectively.

Using these expressions, the corresponding absorption probabilities
were plotted as a function of the energy-parameter $\omega r_H$, for
various values of the number of extra dimensions $n$ and angular momentum
parameter $a_*$. These plots were presented in Section 4. Comparing our
results to numerical results existing in the literature, we found them
to be in very good agreement. The dependence of $|A_{sjm}|^2$ on both
$n$ and $a_*$ was correctly reproduced as well as additional effects
such as the superradiance in the case of gauge bosons. This agreement
extended beyond the  low-$\omega$ and low-$a_*$ regime, even for
intermediate values of
$a_*$ or values of $\omega r_H$ larger than unity. In section 5, we
presented compact, simplified expressions for the absorption probabilities
valid in the limit $\omega \rightarrow 0$. As in the non-rotating case,
the behaviour was found to be different with the absorption probability
for gauge bosons reducing to zero faster than the one for fermions.
This resulted in a different behaviour, in the very low-energy regime,
of the corresponding total absorption cross-sections: while the one for
gauge bosons goes always to zero, the one for fermions assumes a
non-zero, constant value that depends on both $a_*$ and $n$. The
extended validity of our results allowed us to compute the dependence
of the total cross-section over the whole energy regime. We were then
able to derive the high-energy asymptotic values, that turned out
to be the same for fermions and gauge bosons and to coincide with the
one for scalar fields computed in a previous work of ours via the use
of the geometrical optics limit.

Finally, in section 6, we computed the energy emission rates for Hawking
radiation by a rotating, higher-dimensional black hole on the brane in
the form of fermions and gauge bosons. These were also found to be
in remarkable agreement with the exact numerical results in the
low-energy and low-angular-momentum regime, and to lead also to
fairly accurate results beyond those limits.

In conclusion, our study shows that analytical treatment of the
propagation of fields with spin $s$ on a brane embedded in a
higher-dimensional, rotating black hole background is indeed
possible, and that the derived results, although they cannot
offer the level of accuracy and completeness of the exact numerical
ones, can certainly reproduce quite accurately the dependence of the
various quantities on the topological parameters of spacetime
as well as a good estimate of their actual magnitudes.

\bigskip

{\bf Acknowledgments.} S.C and O.E. acknowledge PPARC and I.K.Y.
fellowships, respectively. The work of P.K. is funded by the UK
PPARC Research Grant PPA/A/S/ 2002/00350. K.T. and P.K.
acknowledge participation in the RTN networks
UNIVERSENET-MRTN-CT-2006035863-1 and MRTN-CT-2004-503369. This
research was co-funded by the European Union in the framework of
the Program $\Pi Y\Theta A\Gamma O PA\Sigma-II$ of the
{\textit{``Operational Program for Education and Initial
Vocational Training"}} ($E\Pi EAEK$) of the 3rd Community Support
Framework of the Hellenic Ministry of Education, funded by $25\%$
from national sources and by $75\%$ from the European Social Fund
(ESF).

\

\begin{thebibliography}{99}

\bibitem{ADD} N.~Arkani-Hamed, S.~Dimopoulos and G.~R.~Dvali,
{\it Phys.\ Lett.}\ B {\bf 429}, 263 (1998) [hep-ph/9803315];
{\it Phys.\ Rev.}\ D {\bf 59}, 086004 (1999) [hep-ph/9807344];
I.~Antoniadis, N.~Arkani-Hamed, S.~Dimopoulos and G.~R.~Dvali,
{\it Phys.\ Lett.}\ B {\bf 436}, 257 (1998) [hep-ph/9804398].


\bibitem{RS} L. Randall and R. Sundrum, {\textit{Phys. Rev. Lett.}} $\bf{83}$ (1999) 3370; {\textit{Phys. Rev. Lett.}} $\bf{83}$ (1999) 4690.

\bibitem{creation} T.~Banks and W.~Fischler,
hep-th/9906038;\\ 
D.~M.~Eardley and S.~B.~Giddings,
{\it Phys. Rev.} {\bf D66}, 044011 (2002) [gr-qc/0201034];\\
H.~Yoshino and Y.~Nambu,
{\it Phys. Rev.} {\bf D66}, 065004 (2002) [gr-qc/0204060];\\
{\it Phys.\ Rev.} {\bf D67}, 024009 (2003) [gr-qc/0209003];\\
E.~Kohlprath and G.~Veneziano,
{\it JHEP} {\bf 0206}, 057 (2002) [gr-qc/0203093];\\ 
V.~Cardoso, O.~J.~C.~Dias and J.~P.~S.~Lemos,
{\it Phys.\ Rev.}\ D {\bf 67}, 064026 (2003)[hep-th/0212168];\\
E.~Berti, M.~Cavaglia and L.~Gualtieri,
{\it Phys. Rev.} {\bf D69}, 124011 (2004) [hep-th/ 0309203];\\
V.~S.~Rychkov,
{\it Phys.\ Rev.}\ D {\bf 70}, 044003 (2004) [hep-ph/0401116];\\
S.~B.~Giddings and V.~S.~Rychkov, 
{\it Phys.\ Rev.}\ D {\bf 70}, 104026 (2004) [hep-th/0409131];\\
O.~I.~Vasilenko, 
hep-th/0305067;\\ 
H.~Yoshino and V.~S.~Rychkov,
{\it Phys.\ Rev.} D {\bf 71} (2005) 104028 [hep-th/0503171];\\
D.~C.~Dai, G.~D.~Starkman and D.~Stojkovic,
{\it Phys.\ Rev.} {\bf 73} (2006) 104037 [hep-ph/0605085];\\
H.~Yoshino and R.~B.~Mann,
{\it Phys.\ Rev.} {\bf 74} (2006) {044003} [gr-qc/0605131];\\
H.~Yoshino, T.~Shiromizu and M.~Shibata,
{\it Phys.\ Rev.} D {\bf 74}, 124022 (2006)[gr-qc/0610110].


\bibitem{colliders} S.~B.~Giddings and S.~Thomas,
{\it Phys.\ Rev.}\ D {\bf 65}, 056010 (2002) [hep-ph/0106219];\\
S.~Dimopoulos and G.~Landsberg, 
{\it Phys.\ Rev.\ Lett.}\  {\bf 87}, 161602 (2001) [hep-ph/0106295];\\
S.~Dimopoulos and R.~Emparan, 
{\it Phys.\ Lett.}\ B {\bf 526}, 393 (2002) [hep-ph/0108060];\\
S.~Hossenfelder, S.~Hofmann, M.~Bleicher and H.~Stocker,
{\it Phys. Rev.}\ D {\bf 66}, 101502 (2002) [hep-ph/0109085];\\
K.~Cheung, 
{\it Phys. Rev. Lett.} {\bf 88}, 221602 (2002) [hep-ph/0110163]; \\
R.~Casadio and B.~Harms,
{\it Int. J. Mod. Phys.} \ A {\bf 17}, 4635 (2002) [hep-ph/0110255];\\
S.~C.~Park and H.~S.~Song, 
{\it J.\ Korean Phys.\ Soc.}  {\bf 43}, 30 (2003) [hep-ph/0111069];\\
G.~Landsberg, 
{\it Phys. Rev. Lett.} {\bf 88}, 181801 (2002) [hep-ph/0112061];\\
G.~F.~Giudice, R.~Rattazzi and J.~D.~Wells,
{\it Nucl.\ Phys.}\ B {\bf 630}, 293 (2002) [hep-ph/0112161];\\
E.~J.~Ahn, M.~Cavaglia and A.~V.~Olinto, 
{\it Phys. Lett.}\ B  {\bf 551}, 1 (2003)[hep-th/0201042];\\
T.~G.~Rizzo,
{\it JHEP} {\bf 0202}, 011 (2002) [hep-ph/0201228]; {\it JHEP}
{\bf 0501}, 025 (2005) [hep-ph/0412087];
hep-ph/0611224;\\
A.~V.~Kotwal and C.~Hays,
{\it Phys. Rev.}\ D {\bf 66}, 116005 (2002) [hep-ph/0206055];\\
A.~Chamblin and G.~C.~Nayak,
{\it Phys. Rev.}\ D {\bf 66}, 091901 (2002) [hep-ph/0206060];\\
T.~Han, G.~D.~Kribs and B.~McElrath,
{\it Phys. Rev. Lett.} {\bf 90}, 031601 (2003) [hep-ph/0207003];\\
I.~Mocioiu, Y.~Nara and I.~Sarcevic,
{\it Phys. Lett.}\ B {\bf 557}, 87 (2003) [hep-ph/0310073];\\
M.~Cavaglia, S.~Das and R.~Maartens,
{\it Class. Quant. Grav.} {\bf 20}, L205 (2003) [hep-ph/0305223];\\
D.~Stojkovic,
{\it Phys. Rev. Lett.} {\bf 94}, 011603 (2005) [hep-ph/0409124];\\
S.~Hossenfelder,
{\it Mod.\ Phys.\ Lett.}\ A {\bf 19}, 2727 (2004) [hep-ph/0410122];\\
C.~M.~Harris, M.~J.~Palmer, M.~A.~Parker, P.~Richardson,
A.~Sabetfakhri and B.~R.~Webber,
{\it JHEP} {\bf 0505}, 053 (2005) [hep-ph/0411022];\\
G.~C.~Nayak and J.~Smith,
{\it Phys.\ Rev.} D {\bf 74}, 014007 (2006) [hep-ph/0602129];\\
H.~Stocker,
hep-ph/0605062;\\
L.~Lonnblad and M.~Sjodahl,
{\it JHEP} {\bf 10} (2006) 088 [hep-ph/0608210];\\ 
M.~Cavaglia, R.~Godang, L.~Cremaldi and D.~Summers,
hep-ph/0609001;\\
G.~L.~Alberghi, R.~Casadio and A.~Tronconi,
J.\ Phys.\ G {\bf 34}, 767 (2007) [hep-ph/0611009];\\
M.~Cavaglia, R.~Godang, L.~M.~Cremaldi and D.~J.~Summers,
JHEP {\bf 0706}, 055 (2007) [arXiv:0707.0317 [hep-ph]].
D.~M.~Gingrich,
arXiv:0706.0623 [hep-ph].


\bibitem{cosmic} A.~Goyal, A.~Gupta and N.~Mahajan,
{\it Phys.\ Rev.\ } D {\bf 63}, 043003 (2001) [hep-ph/0005030];\\
J.~L.~Feng and A.~D.~Shapere, 
{\it Phys. Rev. Lett.} {\bf 88}, 021303 (2002) [hep-ph/0109106];\\
L.~Anchordoqui and H.~Goldberg,
{\it Phys.\ Rev.\ }  D {\bf 65}, 047502 (2002) [hep-ph/\-01\-09\-242];\\
R.~Emparan, M.~Masip and R.~Rattazzi,
{\it Phys.\ Rev.\ } D {\bf 65}, 064023 (2002) [hep-ph/ 0109287];\\
L.~A.~Anchordoqui, J.~L.~Feng, H.~Goldberg and A.~D.~Shapere,
{\it Phys.\ Rev.}\ D {\bf 65}, 124027 (2002) [hep-ph/0112247];
{\it Phys.\ Rev.\ }\ D {\bf 68},
104025 (2003) [hep-ph/0307228];\\ 
Y.~Uehara,
{\it Prog.\ Theor.\ Phys.}\  {\bf 107}, 621 (2002) [hep-ph/0110382];\\
J.~Alvarez-Muniz, J.~L.~Feng, F.~Halzen, T.~Han and D.~Hooper,
{\it Phys.\ Rev.}\ D {\bf 65}, 124015 (2002) [hep-ph/0202081];\\
A.~Ringwald and H.~Tu,
{\it Phys.\ Lett.}\ B {\bf 525}, 135 (2002) [hep-ph/0111042];\\
M.~Kowalski, A.~Ringwald and H.~Tu,
{\it Phys.\ Lett.}\ B {\bf 529}, 1 (2002) [hep-ph/0111042];\\
E.~J.~Ahn, M.~Ave, M.~Cavaglia and A.~V.~Olinto,
{\it Phys. Rev.} D {\bf 68}, 043004 (2003) [hep-ph/0306008];\\
E.~J.~Ahn, M.~Cavaglia and A.~V.~Olinto,
{\it Astropart.\ Phys.}  {\bf 22}, 377 (2005) [hep-ph/0312249];\\
T.~Han and D.~Hooper,
{\it New J.\ Phys.}  {\bf 6} (2004) 150 [hep-ph/0408348];\\
A.~Cafarella, C.~Coriano and T.~N.~Tomaras,
{\it JHEP} {\bf 0506}, 065 (2005) [hep-ph/0410358];\\
D.~Stojkovic and G.~D.~Starkman,
{\it Phys.\ Rev.\ Lett.} {\bf 96}, 041303 (2006) [hep-ph/0505112];\\
A.~Barrau, C.~Feron and J.~Grain,
{\it Astrophys.\ J.}  {\bf 630}, 1015 (2005) [astro-ph/0505436];\\
L.~Anchordoqui, T.~Han, D.~Hooper and S.~Sarkar,
{\it Astropart.\ Phys.}  {\bf 25}, 14 (2006) [hep-ph/0508312];\\
E.~J.~Ahn and M.~Cavaglia, {\it Phys.\ Rev.} D {\bf 73}, 042002
(2006) [hep-ph/0511159];
L.~A.~Anchordoqui, M.~M.~Glenz and L.~Parker,
{\it Phys.\ Rev.} D {\bf 75}, 024011 (2007) [hep-ph/0610359].

\bibitem{ADMR}
P.~C.~Argyres, S.~Dimopoulos and J.~March-Russell,
{\it Phys.\ Lett.}\ B {\bf 441}, 96 (1998) [hep-th/9808138].

\bibitem{Kanti2004}
  P.~Kanti,
  Int.\ J.\ Mod.\ Phys.\  A {\bf 19} (2004) 4899
  [arXiv:hep-ph/0402168].


\bibitem{reviews} M.~Cavaglia,
{\it Int. J. Mod. Phys.} {\bf A18}, 1843 (2003) [hep-ph/0210296];
G.~Landsberg, 
{\it Eur. Phys. J.} {\bf C33}, S927 (2004) [hep-ex/0310034];
K.~Cheung, hep-ph/0409028;
S.~Hossenfelder, hep-ph/0412265;
C.~M.~Harris, hep-ph/0502005;
A.~S.~Majumdar and N.~Mukherjee, {\it Int. J. Mod. Phys.}\ D {\bf 14},
1095 (2005) [astro-ph/0503473];
A.~Casanova and E.~Spallucci,
{\it Class.\ Quant.\ Grav.}  {\bf 23}, R45 (2006) [hep-ph/0512063].


\bibitem{hawking} S.~W.~Hawking,
{\it Commun.\ Math.\ Phys.}\  {\bf 43}, 199 (1975).


\bibitem{kmr1}
P.~Kanti and J.~March-Russell,
{\it Phys.\ Rev.}\ D {\bf 66}, 024023 (2002) [hep-ph/0203223];
{\it Phys.\ Rev.}\ D {\bf 67}, 104019 (2003) [hep-ph/0212199].


\bibitem{FS}
V.~P.~Frolov and D.~Stojkovic,
{\it Phys.\ Rev.}\ D {\bf 66}, 084002 (2002) [hep-th/0206046];
{\it Phys.\ Rev.}\ D {\bf 67}, 084004 (2003) [gr-qc/0211055].

\bibitem{HK1} C.~M.~Harris and P.~Kanti,
{\it JHEP} {\bf 0310}, 014 (2003) [hep-ph/0309054].

\bibitem{Barrau} A.~Barrau, J.~Grain and S.~O.~Alexeyev,
{\it Phys. Lett.} B {\bf 584}, 114 (2004) [hep-ph/0311238];
J.~Grain, A.~Barrau and P.~Kanti,
{\it Phys.\ Rev.} D {\bf 72}, 104016 (2005)
[hep-th/\-05\-09\-128];
T.~G.~Rizzo,
{\it Class.\ Quant.\ Grav.}  {\bf 23}, 4263 (2006)
[hep-ph/\-06\-01\-029]; hep-ph/0603242.

\bibitem{Jung}
E.~l.~Jung, S.~H.~Kim and D.~K.~Park,
{\it Phys.\ Lett.} B {\bf 586} (2004) 390 [hep-th/0311036];
{\it JHEP} {\bf 0409} (2004) 005 [hep-th/0406117];
{\it Phys.\ Lett.}\ B {\bf 602} (2004) 105 [hep-th/0409145];
{\it Phys.\ Lett.} B {\bf 614} (2005) 78 [hep-th/0503027];
E.~Jung and D.~K.~Park,
{\it Nucl.\ Phys.}\ B {\bf 717} (2005) 272 [hep-th/0502002];
hep-th/0506204.

\bibitem{BGK}
P.~Kanti, J.~Grain and A.~Barrau,
{\it Phys.\ Rev.} D {\bf 71} (2005) 104002 [hep-th/0501148].

\bibitem{Naylor}
  A.~S.~Cornell, W.~Naylor and M.~Sasaki,
  JHEP {\bf 0602} (2006) 012 [hep-th/0510009].

\bibitem{Park}
  D.~K.~Park,
  Class.\ Quant.\ Grav.\  {\bf 23} (2006) 4101 [hep-th/0512021].


\bibitem{Cardoso}
  V.~Cardoso, M.~Cavaglia and L.~Gualtieri,
  Phys.\ Rev.\ Lett.\  {\bf 96}, 071301 (2006)
  [Erratum-ibid.\  {\bf 96}, 219902 (2006)] [hep-th/0512002];
  JHEP {\bf 0602}, 021 (2006) [hep-th/0512116].


\bibitem{Creek}
  S.~Creek, O.~Efthimiou, P.~Kanti and K.~Tamvakis,
  Phys.\ Lett.\ B {\bf 635} (2006) 39 [hep-th/0601126];
  O.~Efthimiou,
  hep-th/0609144.

\bibitem{Dai}
D.~C.~Dai, N.~Kaloper, G.~D.~Starkman and D.~Stojkovic,
hep-th/0611184.

\bibitem{Liu}
L.~Liu, B.~Wang and G.~Yang,
hep-th/0701166. 

\bibitem{Chen}
S.~Chen, B.~Wang and R.~K.~Su,
Phys.\ Lett.\  B {\bf 647}, 282 (2007)
[arXiv:hep-th/0701209].


\bibitem{Frolov2}
V.~P.~Frolov and D.~Stojkovic,
{\it Phys.\ Rev.}\ D {\bf 67}, 084004 (2003) [gr-qc/0211055].

\bibitem{IOP1} D.~Ida, K.~y.~Oda and S.~C.~Park,
Phys.\ Rev.\ D {\bf 67}, 064025 (2003) [Erratum-ibid.\ D {\bf 69},
049901 (2004)] [hep-th/0212108].

\bibitem{HK2}
C.~M.~Harris and P.~Kanti,
{\it Phys.\ Lett.} B {\bf 633}, 106 (2006) [hep-th/0503010].

\bibitem{IOP-proc} D.~Ida, K.~y.~Oda and S.~C.~Park,
hep-ph/0501210.

\bibitem{DHKW}
G.~Duffy, C.~Harris, P.~Kanti and E.~Winstanley, {\it JHEP} {\bf
{0509}}, 049 (2005) [hep-th/0507274].

\bibitem{CKW}
M.~Casals, P.~Kanti and E.~Winstanley,
{\it JHEP} {\bf 0602}, 051 (2006) [hep-th/0511163].

\bibitem{CDKW} M.~Casals, S.~R.~Dolan, P.~Kanti and E.~Winstanley,
hep-th/0608193.

\bibitem{IOP2}
D.~Ida, K.~y.~Oda and S.~C.~Park,
{\it Phys.\ Rev.} D {\bf 71}, 124039 (2005) [hep-th/0503052];
{\it Phys.\ Rev.} D {\bf 73}, 124022 (2006) [hep-th/0602188].

\bibitem{Jung-super} E.~Jung, S.~Kim and D.~K.~Park,
{\it Phys.\ Lett.} B {\bf 615}, 273 (2005) [hep-th/0503163];
{\it Phys.\ Lett.} B {\bf 619}, 347 (2005) [hep-th/0504139].

\bibitem{rot-bulk} H.~Nomura, S.~Yoshida, M.~Tanabe and K.~i.~Maeda,
{\it Prog.\ Theor.\ Phys.}  {\bf 114}, 707 (2005)
[hep-th/0502179].

\bibitem{Jung-rot} E.~Jung and D.~K.~Park,
{\it Nucl.\ Phys.} B {\bf 731}, 171 (2005) [hep-th/0506204].

\bibitem{CEKT2}
  S.~Creek, O.~Efthimiou, P.~Kanti and K.~Tamvakis,
  Phys.\ Rev.\  D {\bf 75} (2007) 084043
  [arXiv:hep-th/0701288].

\bibitem{MP} R.~C.~Myers and M.~J.~Perry,
{\it Annals Phys.}\  {\bf 172}, 304 (1986).

\bibitem{NP}
E.~Newman and R.~Penrose,
J.\ Math.\ Phys.\  {\bf 3}, 566 (1962).

\bibitem{Chandrasekhar}
S.~Chandrasekhar, {\it {The Mathematical Theory Of Black Holes}}
(Oxford University Press, Oxford, 1992).

\bibitem{press1} W.~H.~Press and S.~A.~Teukolsky,
Astrophys.\ J.\  {\bf 185}, 649 (1973). 

\bibitem{staro} A.~A.~Starobinskii and S.~M.~Churilov,
Zh.\ Eksp.\ Teor.\ Fiz.\ {\bf {65}}, 3 (1973).

\bibitem{fackerell}
E.~D.~Fackerell and R.~G.~Grossman,
J.\ Math.\ Phys.\ {\bf 18}, 1849 (1977).

\bibitem{breuer}
R.~A.~Breuer,
{\it {Gravitational Perturbation Theory and Synchrotron Radiation}},
Lecture Notes in Physics Vol. 44
(Springer, Berlin, 1975).

\bibitem{brw}
R.~Breuer, M.~Ryan Jr. and S.~Waller,
Proc.\ R.\ Soc.\ London A {\bf 358}, 71 (1977).

\bibitem{Casals:2004zq} M.~Casals and A.~C.~Ottewill,
Phys.\ Rev.\ D {\bf 71}, 064025 (2005) [gr-qc/0409012].

\bibitem{churilov} A.~A.~Starobinskii and S.~M.~Churilov,
{\it Sov. Phys.-JETP} {\bf 38}, 1 (1974).

\bibitem{Seidel} E.~Seidel,
Class.\ Quant.\ Grav.\  {\bf 6}, 1057 (1989).

\bibitem{Abramowitz} M. Abramowitz and I. Stegun, {\it Handbook of Mathematical
Functions} (Academic, New York, 1996).

\bibitem{Teukolsky:1973ha} S.~A.~Teukolsky,
Astrophys.\ J.\  {\bf 185}, 635 (1973). 

\bibitem{Unruh}
W.~G.~Unruh,
Phys.\ Rev.\  D {\bf 10}, 3194 (1974);
Phys.\ Rev.\  D {\bf 14}, 3251 (1976).

\bibitem{Guven}
R. G\"uven, Phys.\ Rev.\  D {\bf 16}, 1706 (1977).

\bibitem{Leahy}
D.~A.~Leahy and W.~G.~Unruh,
Phys.\ Rev.\  D {\bf 19}, 3509 (1979).

\bibitem{Detweiler} S.L.~Detweiler,
Proc.\ Roy.\ Soc.\ Lond.\  A {\bf 349}, 217 (1976)

\bibitem{CO} M.~Casals and A.~C.~Ottewill,
Phys.\ Rev.\ D {\bf 71}, 124016 (2005) [gr-qc/0501005].

\bibitem{super}
Y.B. Zel'dovich, JETP Lett. {\bf 14}, 180 (1971).

\bibitem{cross-section} S.~S.~Gubser, I.~R.~Klebanov and A.~A.~Tseytlin,
{\it Nucl. Phys.} {\bf B499}, 217 (1997) [hep-th/9703040];\\
S.~D.~Mathur,
{\it Nucl.\ Phys.\ B} {\bf 514}, 204 (1998) [hep-th/9704156];\\
S.~S.~Gubser, 
{\it Phys.\ Rev.\ D} {\bf 56}, 4984 (1997) [hep-th/9704195].

\bibitem{Emparan}
R.~Emparan, G.~T.~Horowitz and R.~C.~Myers,
{\it Phys.\ Rev.\ Lett.}  {\bf 85}, 499 (2000)
[hep-th/0003118].

\end{thebibliography}
\end{document}